\documentclass[epj]{svjour}
\usepackage{amssymb}    
\usepackage{amsbsy}   
\usepackage{graphicx}

\providecommand{\sta}{\mathrm{sta}}
\providecommand{\tri}{\mathrm{tri}}
\renewcommand{\phi}{\varphi}
\renewcommand{\d}{\mathrm{d}}
\newcommand{\ben}{\begin{equation}}
\newcommand{\een}{\end{equation}}
\newcommand{\bea}{\begin{eqnarray}}
\newcommand{\eea}{\end{eqnarray}}

\begin{document}
%
\title{Decorated vertices with 3-edged cells in 2D foams: exact solutions and properties.}
\author{M. Mancini\inst{1}\thanks{\email{marco.mancini@univ-rennes1.fr}} 
\and C. Oguey\inst{2}
}
\institute{GMCM, CNRS UMR 6626, Universit\'e de Rennes I, 35042 Rennes cedex, France. \and
LPTM, CNRS UMR 8089, Universit\'e de Cergy-Pontoise, 95031 Cergy-Pontoise, France.}

%
%
%
\date{v5.3 \today / Received: date/ Revised version: date}
%
\abstract{
The energy, area and excess energy of a decorated vertex in a 2D foam are
calculated. The general shape of the vertex and its decoration are described analytically by a reference pattern mapped by a parametric Moebius transformation. A single  parameter of control allows to describe, in a common framework, different types of decorations, by liquid triangles or 3-sided bubbles, and other non-conventional cells. A solution is proposed to explain the stability threshold in the flower problem.
\PACS{
     {83.80.Iz}{Emulsions and foams}\and     
     {82.70.Rr}{Aerosols and foams}  \and
      {82.70.Kj}{Emulsions and suspensions}  \and
      {68.03.Hj}{Gas-liquid and vacuum-liquid interfaces: Structure, measurements and simulations}
     } 
} 
\maketitle
%

\section{Introduction} \label{secIntro}

Since J. Plateau \cite{plateau}, 2D foams have been extensively studied, experimentally and theoretically, because they are simpler than three-dimensional systems \cite{WeaireHut}.

In the dry model of 2D soap foams, the gas is assumed incompressible and the liquid fraction is assumed vanishing. Although the dry model correctly describes some aspects of foam physics like energy minima, equilibrium configurations, {\it etc} \cite{WeaireHut}, in many other cases, the presence of the liquid needs to be taken into account, both theoretically \cite{ManciniO} and to match  experimental observations \cite{StavGlazil89,Cox_flower_03,vaz-cox_04}.

The more realistic model of foams, including a small liquid fraction at the vertices, is related to the dry model by Weaire's decoration theorem. The decoration theorem \cite{WeaireBolI} state that, in 2D foams, the films connecting to a Plateau border with vanishing disjoining pressure, if continued inside it, all intersect at a single point and satisfy the equilibrium conditions (fig. \ref{star-triangle}-a). Conversely, it is always possible to decorate a 3-fold vertex at equilibrium by a Plateau border, that is, to replace the vertex by a (small) triangle of liquid without changing the geometry outside the triangle and still satisfying equilibrium. 

The star-triangle equivalence \cite{ManciniO} extends the decoration theorem to general cellular systems where the surface tensions have arbitrary values; this includes 3-sided bubble in standard foams (constant uniform surface tension) or in non standard foams (where the surface tension takes different values on different films \cite{adlerGamm95,adlerGamm00}), or Plateau borders with non vanishing disjoining pressure. In this context, bubbles, drops and Plateau borders are treated on a common footing as {\it cells}.

The decoration theorem validates the ideal dry model for slightly wet foams as far as equilibrium is concerned \cite{WeaireHut}; the star-triangle equivalence permits, in principle, to take away 3-sided bubbles in searching equilibrium configurations. However, the foam energy, the area of the bubbles, coarsening and most  mechanical properties of the foam change if the decoration is switched on or off.  

In dry foams, since the energy is the product of surface tension by film length, minimisation of line-length at fixed bubbles area completely determines the 2D foam's equilibrium structure. The energy of a progressively strained dry foam increases until two threefold vertices approach one another and undergo a T1 neighbour exchange \cite{princen83,cantat05,holer05}.
The T1 quickly reduces the energy by a finite amount: the energy difference between the configuration with an unstable fourfold vertex and the one with a pair of threefold vertices linked by a film. That is why the location and statistics of T1 events determine the inelastic properties of the foam and how it releases energy under strain or quasi-static flow.

If the foam is not dry, the presence of Plateau borders reduces the film length, and thus favours the switch compared to a dry foam \cite{hutzler_05}. At the switch, two triangular Plateau borders meet, merge \cite{brakke,fortes05apr} and then split to a new configuration involving again two distinct Plateau borders.  
So, in terms of energy  furnished to the foam, the presence of Plateau borders reduces the barrier one needs to overcome to trigger a T1.
Furthermore, it is well known that the yield stress of a foam decreases with the liquid fraction \cite{Mason,Gardiner,SJ-Durian,addad_05}. 

By similar considerations, when 3-sided bubbles are inserted at some vertices of the foam, one can expect an increase of the T1 energy barrier and, therefore, a reduction of the shear plasticity due to edge flips.

In this paper, we calculate the energy of a wet-deco\-rated foam starting from the energy of the dry model.    

Recently, Teixeira and Fortes \cite{Fortes05} gave the equations describing the exact geometry of a general Plateau border with zero disjoining  pressure: the surface tension of the films and of the liquid-gas interfaces are related by $\gamma_{\mathrm{film}}=2\gamma_{\mathrm{border}}$; they calculated the excess energy among other quantities.
The excess energy is the energy difference between a decorated and a dry vertex. 

Using the invariance of 2D foams under Moebius transformations \cite{ManciniEuf04}, we generalise this problem to the cases where the surface tension of the films and that of the decorating triangles are arbitrary, giving the Plateau border and 3-sided bubble problems a unified description. We calculate the excess energy $E$ and all the geometrical quantities as functions of four parameters  which characterise the cluster size and shape. 
Normalising the excess energy by the square root of the triangle area gives a scale invariant quantity $\epsilon_\beta$: the {\it relative excess energy}.

In realistic foams, the relative excess energy of 3-cells is found to be approximately linear as a function of the triangle area. The slope of $\epsilon_\beta$ depends on the form of the decorated films.
According to Lewis' law \cite{Lewis}, the average cell area increases linearly with the number of sides or neighbours: $\langle A_n\rangle \propto n$. So the triangle area $ A_3$ is small on average. As we'll see, the slope  is proportional to the sum of the squared curvatures of the films meeting at the decorated vertex. Being zero in the 3-fold symmetric case, the slope of $\epsilon_\beta$ measures the deviation from perfect symmetry.

In the final part, we apply our analytical results to the flower problem.
In the experiments, a bubble spontaneously gets out \cite{Cox_flower_03} before the critical area is reached where the dry model predicts a spontaneous symmetry reduction \cite{Wea_Cox_Gran_flower_02}.
Following the experimentalists' suggestion that the ejection might be due to the presence of liquid, we solve the problem by including liquid triangles around the vertices and calculating a threshold area for the central bubble at which two vertices get into contact. Our conjecture is that this contact triggers the ejection.

The paper is organised as follows: Section 2 recalls the equilibrium equations for 2D foams with non-constant surface tensions and the star-triangle equivalence. 
Section 3.1 defines the reference structure which is then used, in sec. 3.2 and 3.3, to describe general 3-cells, including Plateau borders, 3-sided bubbles and similar patterns. Section 4 contains the calculation of the main quantities: the energy gap between bare and decorated structures, the area of the 3-cell and the relative excess energy. In section 4.4, the relative excess energy is studied as a function of the pressure of the four bubbles involved. 
Section 5 contains the series expansion of the relative  excess energy with respect to the  3-cell area when the star films are fixed. Finally, in Section 6, we apply our analytical results to the flower problem. 

\section{Equilibrium of 2D foams} \label{secEquil}

To describe various types of 2D foams, we consider here a general 2D cellular system with surface tensions verifying the following equilibrium properties. 

The cells are separated by interfaces obeying Laplace-Young's law:
\begin{equation}\label{ulaplace}
  \Delta P + \gamma k =0\,.
\end{equation}
$\Delta P=P_2-P_1$ is the pressure difference across the interface, $\gamma$ the surface tension and $k$ the curvature of the interface \cite{WeaireHut}.
In static conditions and in absence of applied field, the pressure inside each cell is constant so that the interfaces are arcs of circles\footnote{Even if we admit surface tension varying from edge to edge, we always assume that it remains constant inside every edge.}.

At equilibrium, the interfaces meet three by three at vertices in a manner satisfying Plateau's laws \cite{plateau,WeaireHut} :
\begin{eqnarray}
  \label{uplateaub}&\sum_{j=1}^3\gamma_j \mathbf{t}_j=0\,,\\
  \label{uplateauk}&\sum_{j=1}^3\gamma_j k_j=0\,,
\end{eqnarray}
where  $\mathbf{t}_j$ is the unit vector tangent to the interface at the vertex and where the surface tension, $\gamma_j$ ($\gamma_j>0$), differs from interface to interface \cite{ManciniO,adlerGamm95,adlerGamm00,Moukarzel}.

Equations (\ref{uplateaub}) and (\ref{uplateauk}) imply that the symmetry group of 2D cellular systems at equilibrium is the group of homographies, or Moebius maps, generated by Euclidean similarities and inversion \cite{Moukarzel,Weaire}. Some definitions and basic properties of Moebius maps are recalled in appendix \ref{app_a}. 

Equations (\ref{uplateaub}) and (\ref{uplateauk}) also imply that the centres of curvature of  three circular edges meeting at a vertex are aligned \cite{Moukarzel}.

This common framework describes several situations.
A cell containing liquid is a Plateau border. A cell filled with gas is commonly called a bubble.
The sides of a liquid cell are liquid-gas interfaces, with liquid-gas surface tension $\gamma_{\ell g}$.
The edges separating (gaseous) bubbles are liquid films, containing a negligible amount of liquid compared to Plateau borders. Films have a specific surface tension $\gamma_{\mathrm{film}}$.
In standard dry equilibrated 2D foams, the surface tension is constant over the entire foam ($\gamma_j=\gamma=\gamma_{\mathrm{film}}$). In non-standard dry 2D foam, the surface tension may vary from film to film ($\gamma_j\ne\gamma_{j'}$) \cite{adlerGamm95,adlerGamm00}.
Slightly wet equilibrated 2D foams contain bubbles and a small amount of liquid mainly confined in Plateau borders forming concave triangles around the bubble vertices.  

The energy of a 2D cellular cluster is the sum, over the set of edges $j$, of the edge length, $l_j$, weighted by  surface (line in 2D) tension:
\ben\label{enefoam} E= \sum_j\gamma_j\, l_j\,.\een

\subsection{Star--Triangle Equivalence}\label{secS-T}

A {\it triangle} is a three-sided equilibrated cell. Each vertex is the end of exactly 3 curved edges; two of them are part of the triangle boundary but the third one is external, outside the triangle. If the foam is viewed as a graph, this edge connects the triangle to the rest of the foam. Each triangle has three such connecting edges, or bonds, one at each vertex.

A {\it star} is an equilibrated figure formed by a vertex and the three edges meeting at it.

At equilibrium the pressure, or the area, and the surface tensions fix the triangle geometry. The star--triangle equivalence states the following \cite{ManciniO}:
\begin{itemize}
\item For any triangle, the three external bonds, if extended inside the triangle with the same curvature, all intersect at a common point, that forms a star with the extended edges. 
\item Conversely, given any star, there is a continuous range of areas extending down to 0, and an open set of tension values, such that, for any prescribed values in these ranges, there is a triangle with one vertex on each branch of the star that forms an equilibrated figure once the portion of the star inside the triangle has been removed.
Briefly said, the triangle decorates the star. 
\end{itemize}
\begin{figure}[!ht]\centering
a) \includegraphics[width=.4\columnwidth]{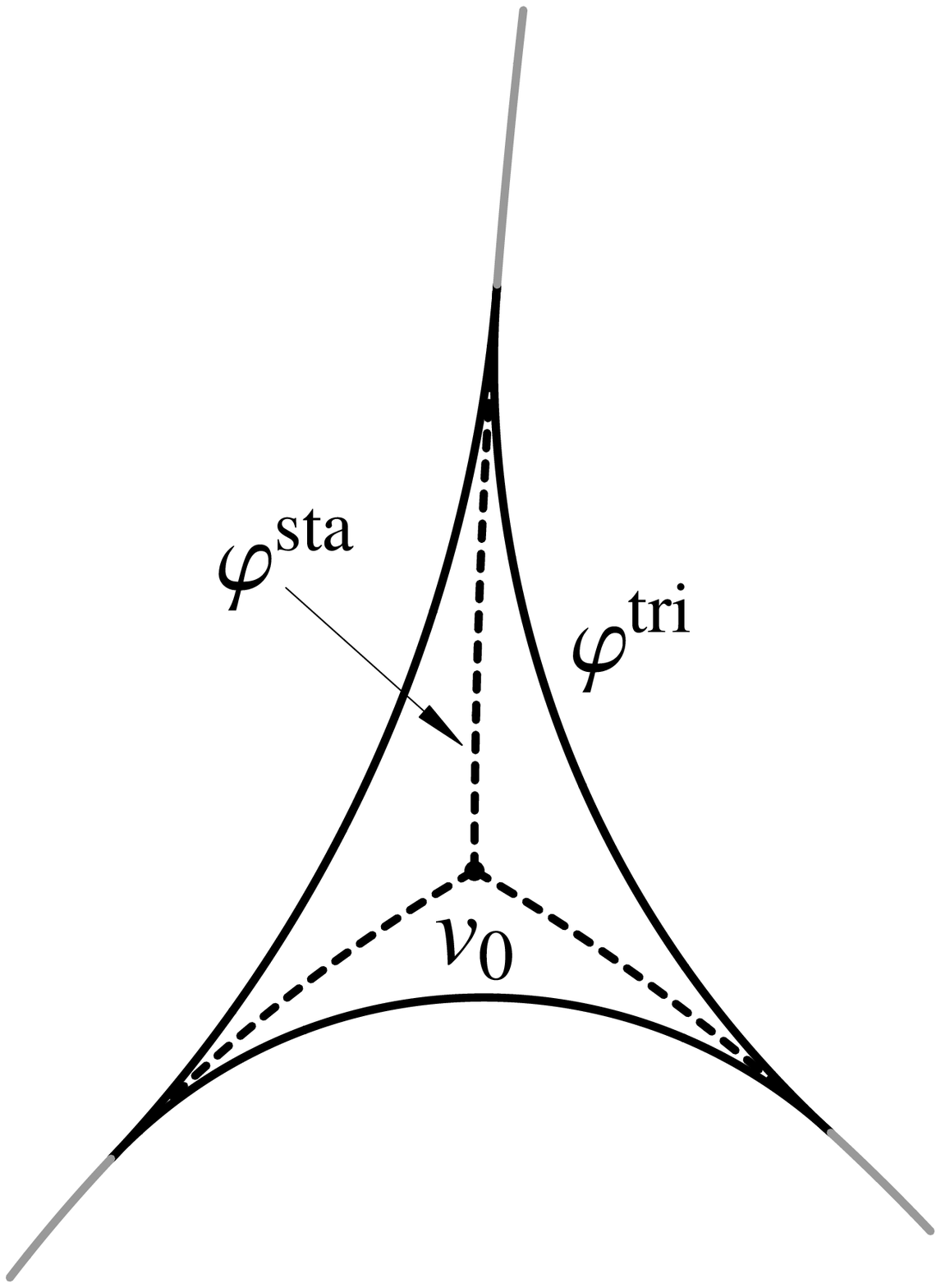}
\quad b) \includegraphics[width=.4\columnwidth]{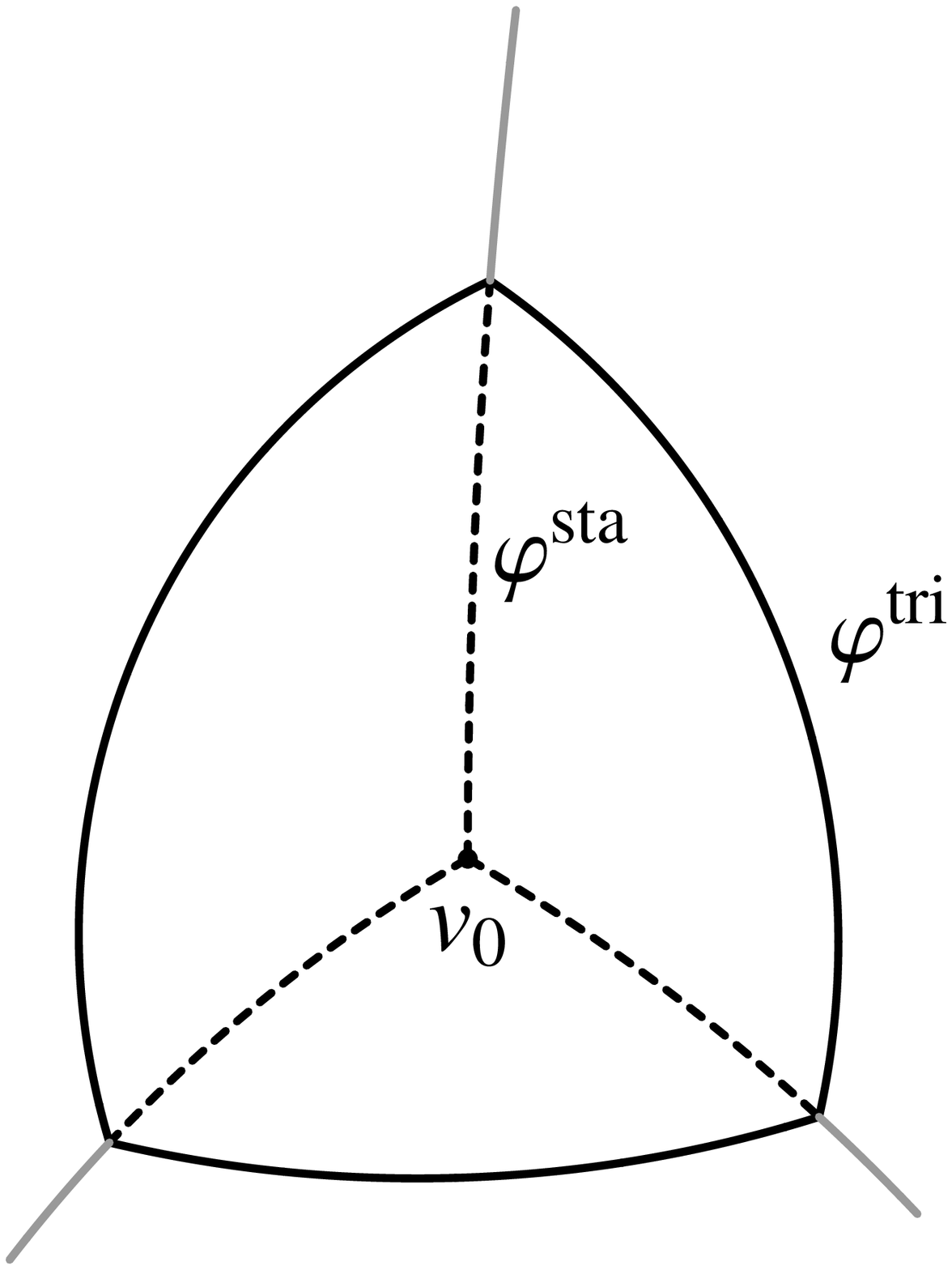}
\caption{Star--triangle equivalence in the case of a Plateau border (a) and a three-sided bubble (b).  The triangle (solid lines) can be replaced by the star (dotted) and, conversely, the star can be decorated by the triangle, in a way preserving equilibrium.  The connecting films (in grey) around the triangle continue, as dotted arcs of circles, inside the triangle to meet at the equilibrated point $v_0$.}
\label{star-triangle}
\end{figure}
According to this equivalence, the star and triangle can replace one another without any modification in the rest of the foam, and in a way preserving all the equilibrium conditions.
To be precise, both figures have the same contact points with the outside foam, namely the triangle vertices; in the substitution, the star branches are limited to the segments inside the triangle. The external parts are kept unchanged.
The star--triangle equivalence was proved in full generality for cellular systems with multiple surface tensions  \cite{ManciniO}. 

When we consider a triangle and its associated star as a whole, a {\it star+triangle}, we will call it {\it 3-cell}. We need the complete figure to evaluate the excess energy, for example. But it should stay clear that, physically, the star and triangle cannot be both present at the same time. The only exceptions will be found in sec. \ref{sec_remark}.

The star--triangle equivalence contains Weaire's decoration theorem \cite{WeaireBolI} by triangular Plateau borders as a particular case (fig. \ref{star-triangle}-a): when the surface tension of the triangle sides is that of a liquid-gas interface, $\gamma_e\equiv\gamma_{\tri}=\gamma_{\ell g}$, while the star has the film surface tension,
$\gamma_i\equiv\gamma_{\sta}=\gamma_{\mathrm{film}}\simeq 2\gamma_{\ell g}$, at zero disjoining pressure.
 
For a standard dry foam, star-triangle equivalence involves a star and a 3-sided bubble with the same tension everywhere: $\gamma_i= \gamma_e=\gamma_{\mathrm{film}}$.

In 3 dimensions, a decoration-bubble theorem holds only for spherical foams \cite{MancioTesi}. When the films are not necessarily spherical, Teixeira and Fortes proposed a modified, approximate, version involving both line and surface tensions \cite{FortTex_DEC05}. 

\section{Conformal description of 3-cells}\label{conf_descr_3_cell}

In this section the description of the 3-cell ({\it star+triangle}) is given. 
First, we construct a reference 3-cell depending on two parameters.
Then, a general 3-cell is obtained by a Moebius transformation.

To fix notations (see fig.\ref{star-triangle}), let $\phi^{\tri}$, of radius $r^{\tri}$, be the  edges of the triangle (decorating edges), and $\phi^{\sta}$, of radius $r^{\sta}$, the  edges of the star (decorated or internal edges). When the star, with vertex $v_0$, is decorated by a triangle, $v_0$ and $\phi^{\sta}$ are virtual. Conversely, when the star is not decorated, the $\phi^{\tri}$ are virtual. 
We will use the convention of signed angles and radii, in a way such that the arc length is always positive \cite{WeaireBolI}.

In this paper, we allow only two different values for $\gamma$:
the surface tension on the triangle boundary sides is $\gamma_e\equiv\gamma_{\tri}$, and that of the external and internal films, on the star edges, is $\gamma_i\equiv\gamma_{\sta}$. These tensions are related by equation (\ref{uplateaub}), applied to a vertex of the triangle:
\ben\label{equivert}
\gamma_i=2\,\gamma_e \cos{\alpha}.
\een
The {\it contact angle} $\alpha$ is the angle between $\phi^{\sta}$ and $\phi^{\tri}$.

If the thickness of the films, $h$, is not negligible, the angle $\alpha$ is related to the disjoining  pressure $\Pi$ \cite{adlerGamm95,adlerGamm00,MancioTesi}:
$$\Pi=\frac{\gamma_i}{h}(\cos{\alpha}-1)\,.$$

The excess energy is the energy gained by decorating the vertex.
It is the  difference of the internal and triangle edge lengths, weighted by the surface tensions \cite{Fortes05}:
\bea\label{excessEdef}
\nonumber E&\equiv&E_{\mathrm{triangle}} - E_{\mathrm{star}}= \gamma_e L^{\tri}-\gamma_i L^{\sta}\\
&=&\gamma_i \left(\frac{L^{\tri}}{2\,\cos \alpha}-L^{\sta}\right).
\eea

The {\it relative excess energy}  is defined as the ratio of the excess energy $E$ over the square root of the triangle area:
\ben\label{def_excess_ene}
\epsilon\equiv \frac{E}{A_{\mathrm{triangle}}^{1/2}}\,.
\een

From now on, we will set $\gamma_i=1$; the surface tension of the internal films is our tension unit.

\subsection{The reference 3-cell}\label{refer_3_cell}
\begin{figure}[!ht]\centering
\includegraphics[width=.8\columnwidth]{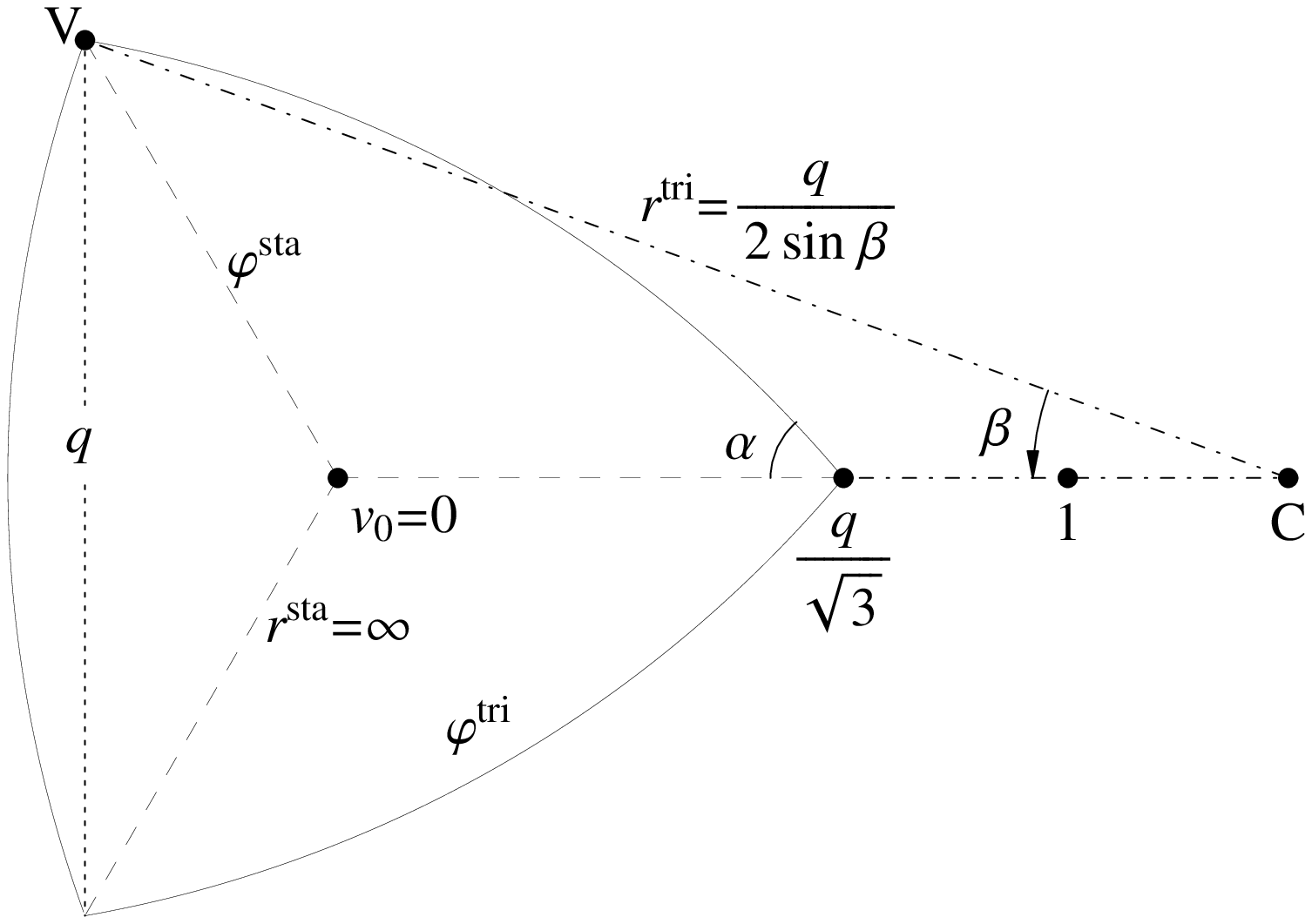}
\includegraphics[width=.9\columnwidth]{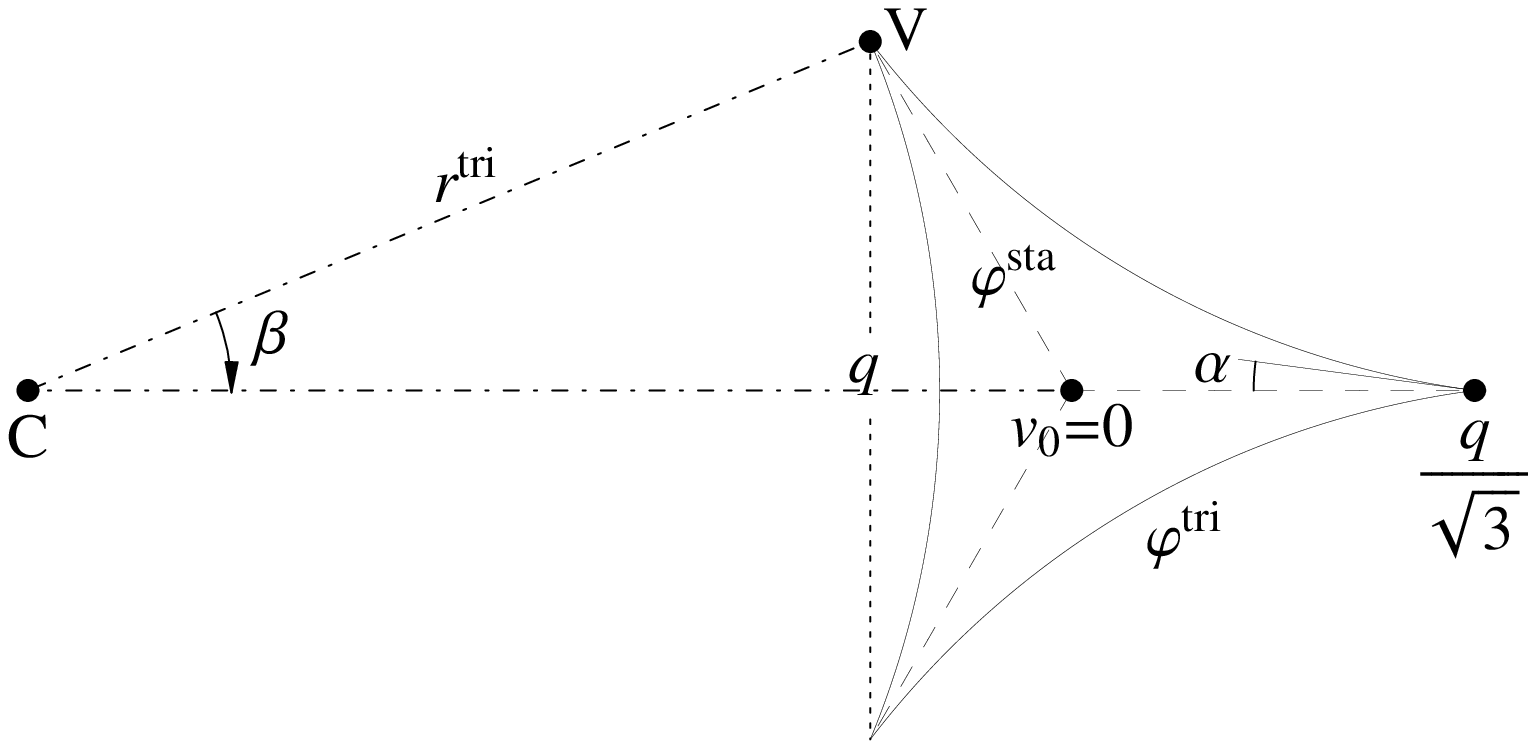}
\caption{The reference, symmetric, 3-cell with parameters $\beta$ and $q$. The star (internal films) is represented by dashed lines.
$VC$ is a radius of the left side. Solving the triangle $VOC$ shows that the angle at $C$ is indeed $\beta=\alpha-\pi/6$. The case of a bubble, $\beta>0$, is on top; the Plateau border case, $\beta<0$, is below.}
\label{symmcell}
\end{figure}

The reference 3-cell $\mathcal{C}(q,\beta)$, of parameters $\beta$ and $q$, is the completely symmetric star+triangle defined in figure \ref{symmcell}.

The parameter $q\ge 0$ is the chord length of the curved sides of the reference triangle.
It specifies the triangle size. 

The parameter $\beta$ is the angle defined by
$$\beta\equiv\alpha-\pi/6\,.$$ 
As conformal means angle preserving, the contact angle $\alpha$, and so $\beta$, are preserved by the Moebius transformation applied to, or from, the reference cell. 
Choosing $\beta$, rather than $\alpha$, helps in treating Plateau borders and 3-sided bubbles in a common way.  In figure (\ref{star-triangle}-a), $\beta$ is equal to $\pi/6$; in fig. (\ref{star-triangle}-b), $\beta=-\pi/6$. 
In the reference cell, the angle subtended by the triangle sides equals $|2\beta|$; this property is not generally preserved by conformal transformations.

When $\beta=\pi/3$, $\gamma_i/\gamma_e=0$ and the sides of the triangle $\mathcal{C}(q,\beta)$ form a circle. This case describes a 3 cells partition of a disc surrounded by a rigid membrane.
So, the interpretation of the triangle as a bubble or a liquid drop is limited to the values $\beta\, \in\,\lbrack-\pi/6,\pi/3\rbrack$.

The edge curves are parametrised by
\bea
\label{fi_int_z}\phi^{\sta}_j(t_1)&=&e^{i\theta_j }\,t_1\,,\qquad \textrm{with}\;t_1\in\lbrack 0,q/\sqrt{3}\rbrack,\\
\label{fi_est_z}
\phi^{\tri}_j(t_2)&=&\frac{q e^{i\theta_j }}{2\sin(\beta)}\,\left(\frac{2}{\sqrt{3}}\cos(\beta+\pi/6)-e^{i\, \beta \,t_2} \right),
\eea
with $t_2\in\lbrack -1,1\rbrack$; $j=1,2,3$ indexes the edges and $\theta_j\equiv 2\pi\,j/3$. 

The curve (\ref{fi_est_z}) depends continuously on $\beta\in \lbrack-\pi/6,\pi/3\rbrack$. Its curvature vanishes, and changes sign, at $\beta=0$. The equilateral triangle is curved, convex, at positive $\beta$; curved, concave, at negative $\beta$; and it has straight edges at $\beta=0$. 
At $\beta=0$ the parametrisation is given by the limit of (\ref{fi_est_z}) as $\beta\to 0$:
\ben\label{phi_ext_lim}
\phi^{\tri}_j(t_2)= -q e^{i\theta_j} \frac{1}{2} \left(3^{-1/2}+i\,t_2\right)\,.
\een

With (\ref{excessEdef}), the excess energy of the reference cell is
\ben\label{excessEbase}
E_\beta(q)=\left( \frac{3\beta}{2\sin{\beta}\cos{(\beta+\frac{\pi}{6})}}-\sqrt{3}\right)q\,.
\een

The triangle area is the sum of the area, $A_\bigtriangleup$, of the rectilinear equilateral triangle based on the same vertices, and $A_\cup$, the signed area of the three lenses around the straight triangle:
\ben\label{areabase}
A_\beta(q)=A_\bigtriangleup+A_{\cup}=
\left(\frac{\sqrt{3}}{4}+\frac{3}{8}\,
\frac{2 \beta-\sin(2 \beta)}{\sin^2\beta}\right)q^2\,.
\een

Finally, the reference relative excess energy is obtained by dividing the excess energy by the area square root: 
\bea\label{epsbase}
\epsilon_\beta^0&=& \frac{E_\beta(q)}{A_\beta(q)^{1/2}}\nonumber\\
&=& \mathrm{sign}(\beta)\frac{\sqrt{3 \beta-2 \sqrt{3}\,\cos{(\beta+\frac{\pi}{6})}\, \sin{\beta}}}{\cos{(\beta+\frac{\pi}{6})}}\,.
\eea
Being dimensionless, $\epsilon_\beta^0$ does not depend on $q$. 
In the particular cases of a Plateau border and of a 
3-sided bubble, the respective $\epsilon_\beta^0$ are
\bea
\nonumber\epsilon_{-\pi/6}^0&=&-\sqrt{\sqrt{3}-\pi/2}\simeq -0.401565\,,\\ 
\nonumber\epsilon_{\pi/6}^0&=&\sqrt{2(\pi-\sqrt{3})}\simeq 1.67901\,.
\eea

As a final remark on the reference, notice that the star is fixed, made of straight edges meeting at $2\pi/3$, the standard Plateau angles. In particular, the star is independent of the parameter $\beta$ and it is affected by $q$ only regarding the length at which its edges are cut. This indicates why, in all subsequent formulae, the quantities concerning only the star do not depend on $\beta$, nor on $q$ except for side length.
 
\subsection{The general 3-cell}\label{gene_3_cell}
Applying a suitable Moebius transformation to the reference 3-cell produces a general 3-cell at equilibrium (the star and triangle both satisfy eqs. (\ref{ulaplace}), (\ref{uplateaub}) and (\ref{uplateauk})). 

Given a star, the decorating triangle is entirely specified by its area $A$ and contact angle $\alpha$; those will be related to the parameters $q$ and $\beta$ of the reference 3-cell.
Now, our transformation must map the (fixed) reference star onto a general star of any possible shape (at equilibrium, of course).
But, in the star, centred at vertex $\tilde{v}^0$, the three (internal) films $i=1,2,3$ have curvature $k_i$. Because of equation (\ref{uplateauk}), only two of the curvatures are independent. So, removing the Euclidean degrees of freedom, we need only a one-complex parameter Moebius transformation to reach all possible star+triangle figures. 

\begin{figure}[!ht]\centering
a)\includegraphics[width=.4\columnwidth]{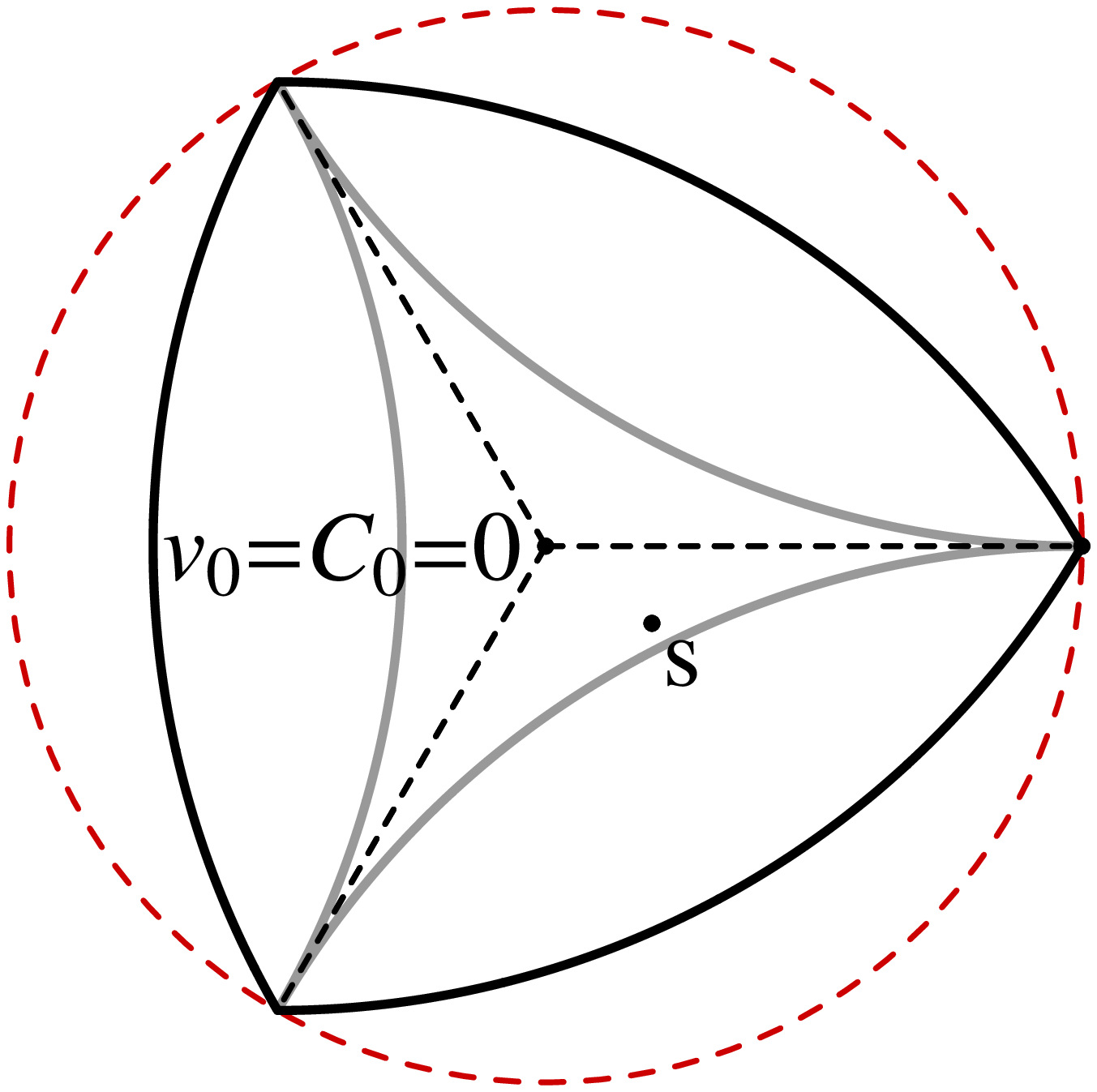}
\qquad b)\includegraphics[width=.4\columnwidth]{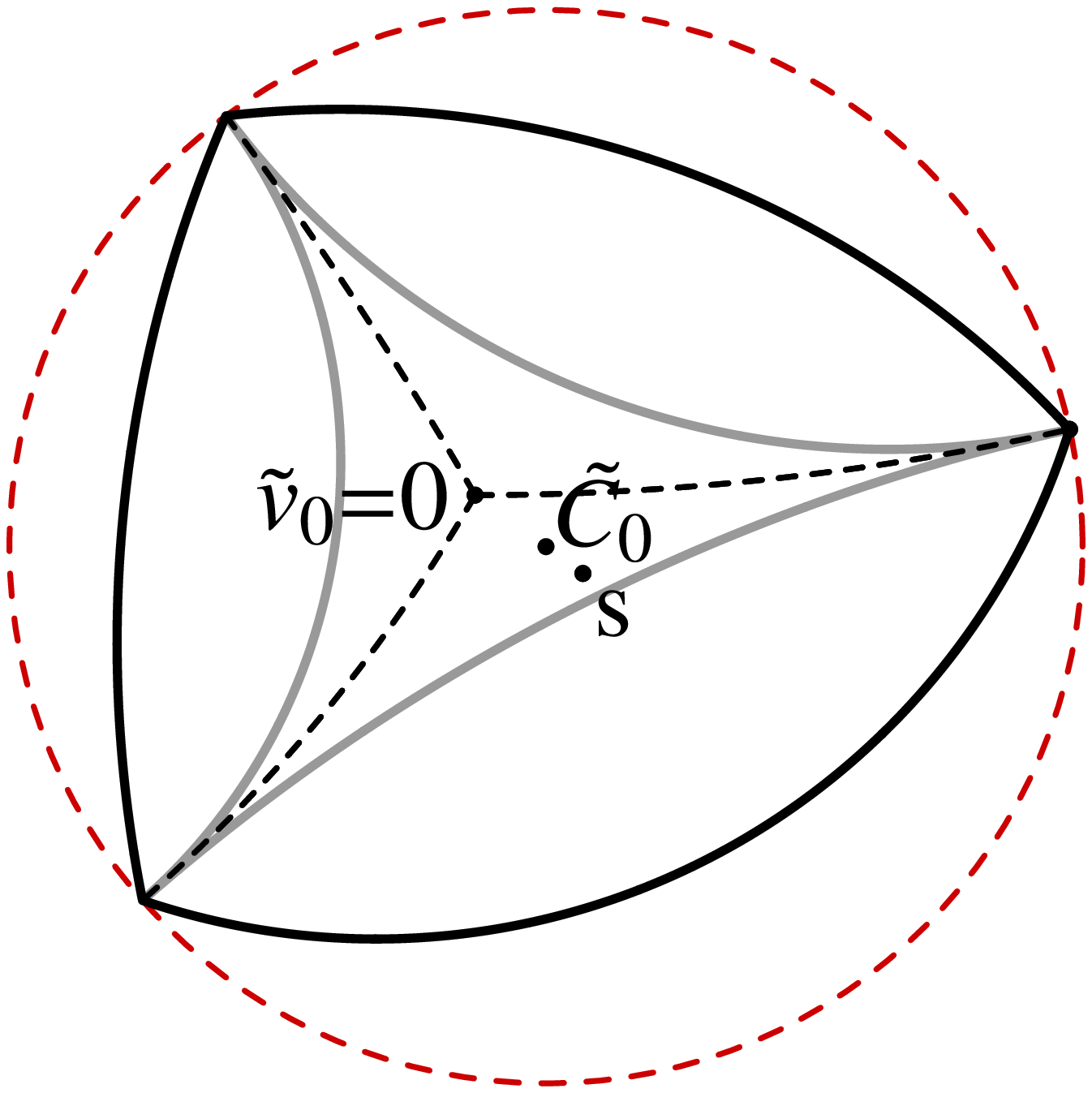}
\caption{a) The reference 3-cell $\mathcal{C}(q,\beta)$, $\beta=\pi/6$ (in black) and $\beta=-\pi/6$ (in grey). b) General 3-cell $\tilde{\mathcal{C}}(s,q,\beta)$, image of a) by $f_s$. The figures are calculated for $q=\sqrt{2}$ and  $s=0.2\exp{(-i\frac{\pi}{5})}$. The circumscribed circle in b) is given by (\ref{cerchio_trasf}).}
\label{MoebImage}
\end{figure}

A general Moebius transformation $f$, given by (\ref{moeb_gen_trans}), has  
six  real parameters, much beyond our needs. We require that the transformation 
\textit{i}) preserves orientation and 
\textit{ii}) maps the interior of the reference 3-cell onto the interior of the general 3-cell. 
In addition, we can fix a length scale, the star vertex position, and one of the film tangents there (by conformality, this fixes all the tangents at $v^0$). Choosing the origin at $v^0$, this means $f(0)=0$ and $f'(0) > 0$. Then the Moebius transformation can be written in the form
\ben\label{moeb_transf}
f_s(z) = (1-|s|^2) \frac{z}{1-\bar{s}\,z}\,,\quad\textrm{with}\ |s|^2=s\,\bar{s}<1\,,
\een
depending on the single complex parameter $s$.
The result is a 4 (real) parameters star+triangle figure: $\widetilde{\mathcal{C}}(s,q,\beta)=f_s(\mathcal C(q,\beta))$. We designate the transformed quantities by a tilde (appendix \ref{app_a}). An example is given in figure \ref{MoebImage}.

The points $z=0$ and $s$ are fixed points of $f_s$. 
The complex function $f_s$ maps the reference circumscribed circle, of centre $C_0=v_0=0$ and radius $r_0=q/\sqrt{3}$ (fig. \ref{MoebImage}), onto the circle of parameters (eq. (\ref{trasf_cer_cer})):
\ben\label{cerchio_trasf}
(\tilde{C}_0,\tilde{r}_0)=
\left(q^2\frac{s(1-|s|^2)}{3-q^2|s|^2},
q\frac{\sqrt{3}(1-|s|^2)}{|3-q^2|s|^2|}\right)\,.
\een

The 3-cell is contained in the disc $(\tilde{C}_0,\tilde{r}_0)$ for $|s|\,q<\sqrt{3}$, while it is outside\footnote{This peculiar situation is possible only when the triangle is curved and has two concave sides.} the disc for $|s|\,q>\sqrt{3}$.

As the equilibrium equations (\ref{uplateaub}) and (\ref{uplateauk}) are left invariant by $f_s$, the image of the reference, either as a star or as a triangle, is in mechanical equilibrium.

In order to get physically meaningful patterns, $q$ must be bounded: $q < q_{\max}(s,\beta)$. An expression for $q_{\max}(s,\beta)$ will be derived in section \ref{subsec_peri}. 
Here, we just give its origin and meaning.
At $q = q_{\max}$, the reference boundary, $\partial\mathcal{C}(q,\beta)$, meets $1/\bar{s}$, the pole of $f_s$; this condition defines $q_{\max}(s,\beta)$. When $q < q_{\max}$, the reference triangle is mapped to a bounded triangle but, when $q > q_{\max}$, the reference interior is mapped to the exterior of the triangle boundary, of infinite area, violating our requirement \textit{ii}), above.

\begin{figure}[!ht]\centering
\includegraphics[width=.95\columnwidth]{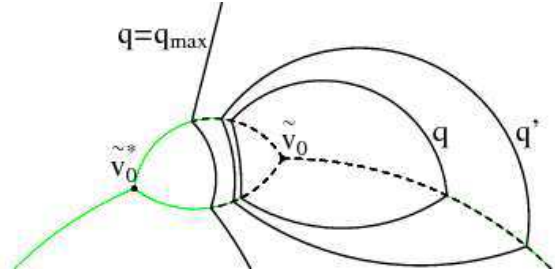}
\caption{Two different 3-cells $\tilde{\mathcal{C}}(s,q,\beta)$ calculated for $\beta=\pi/6$, $q$ and $q'$, with $q'>q$. The continuations of the internal films is plotted in grey (green on-line).}
\label{errelimit}
\end{figure}

The divergence at $q\rightarrow q_{\max}$ is illustrated in figure \ref{errelimit}.
First, recall that any equilibrated star vertex has a {\it conjugate} $v^*$ where the circles supporting the star edges also meet. The position of $v^*$ is the mirror reflection of $v$ through the line of the curvature centres \cite{Moukarzel,MancioTesi}.
 For the reference $\mathcal{C}(q,\beta)$, the conjugate of $v_0=0$ is $v_0^*=\infty$; in $\widetilde{\mathcal{C}}(s,q,\beta)$, $\tilde{v}_0^*$ is given by
\ben
\tilde{v}_0^*=f_s(\infty)= -s\,(|s|^{-2}-1)\,.
\een
By increasing $q$ alone, the triangle vertices move along the star edges, towards infinity in the reference, and so, towards $\tilde{v}_0^*$ under the mapping $f_s$ (figure \ref{errelimit}). When  $q \geq q_{\max}$, the triangle area is infinite and all the edge radii are negative.

In real foams, where the star edges end at neighbour vertices, this divergence is not reached. The route to it is deviated by topological changes.
Indeed, suppose that the area of a 3-sided bubble is increased progressively, for example by injecting gas; much before one sees its area diverging, topological changes will occur, either transforming the triangle into a higher polygon, or breaking films.

\subsection{Extension to special 3-cells}\label{sec_remark}
Although the 3-cell is well defined only for values $-\pi/6\le\beta\le\pi/3$, the equations can be extended, for particular values of the parameters, to cover other physical or mathematical situations. These cases are special because the star and triangle are taken here altogether to form a complete equilibrated cluster, without any outside foam. With contracting tensions, this is only possible for obtuse contact angles, that is, $\beta\geq\pi/3$.
  
\subsubsection{Case $\beta=\pi/3$: triple partition of the disc}\label{remarkcanete} 
As noticed in sec. \ref{refer_3_cell}, the reference $\mathcal C(q, \beta=\pi/3)$ is a disc. $f_s$ maps it to a disc divided into three bubbles by films meeting the outer boundary orthogonally\footnote{$\alpha=\pi/2$ is the angle formed by the soap films with a smooth rigid wall or membrane.}.
Ca\~nete and Ritor\'e \cite{canete04} proved that this type of graph is the unique least-perimeter way of partitioning the unit disk into three regions of prescribed areas.
\begin{figure}[!ht]\centering
\includegraphics[width=.4\columnwidth]{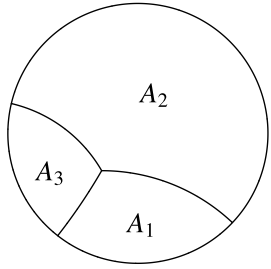}
\caption{For $q=\sqrt{3}$ and $\beta=\pi/3$, varying $s$, the star describes all the least-perimeter partitions of the unit disk into three regions of prescribed areas.}
\label{canete}
\end{figure}

\subsubsection{Case $\beta=\pi/2$: three bubble cluster}\label{remark3clust}
\begin{figure}[!ht]\centering
a)\includegraphics[width=.4\columnwidth]{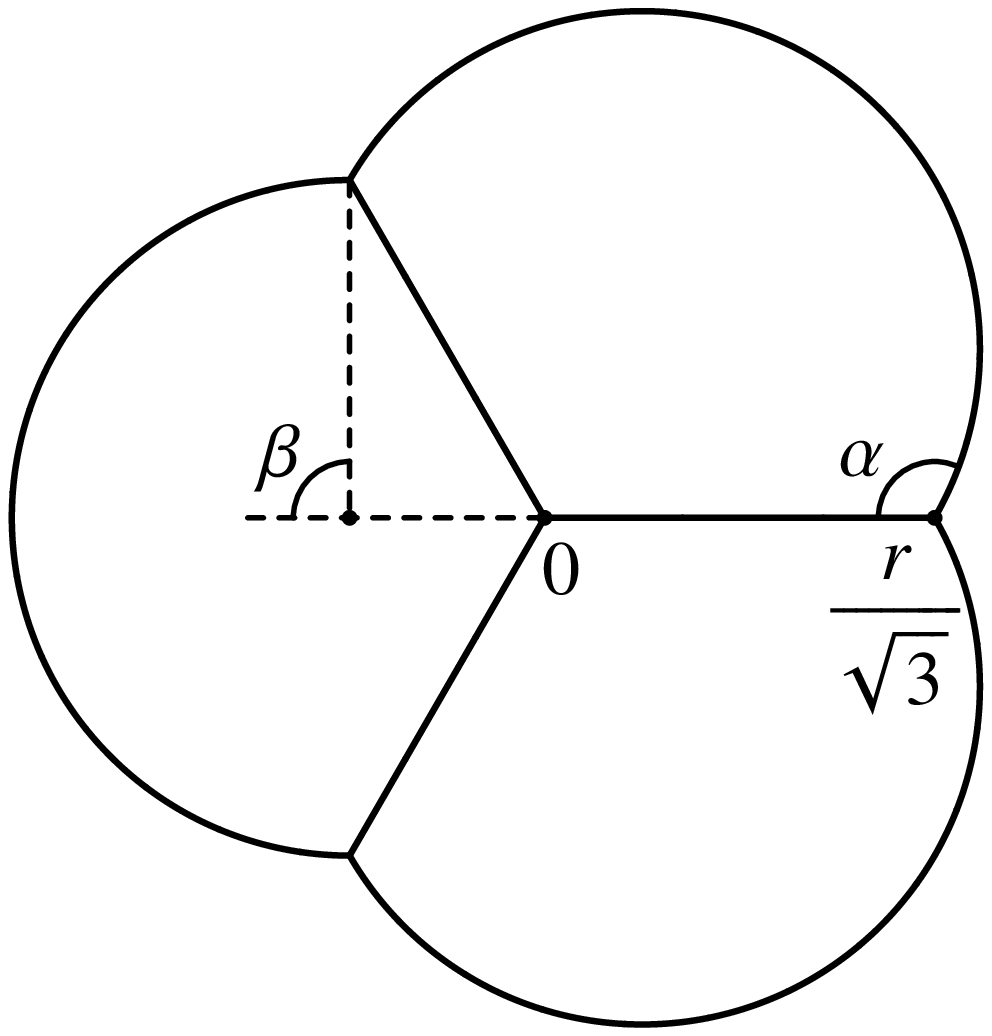}
\qquad
b)\includegraphics[width=.4\columnwidth]{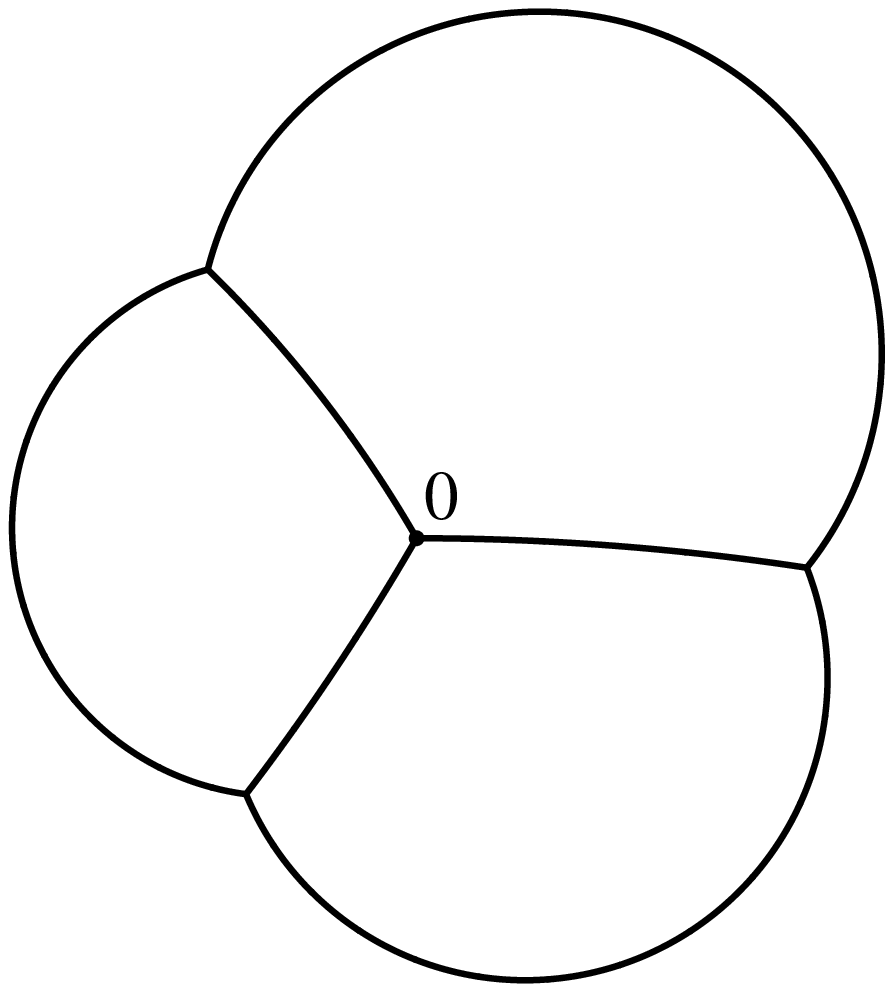}
\caption{a) At $\beta=\pi/2$, the reference describes a cluster of 3 identical bubbles.
b) Three-bubble cluster transformed by $f_s$.}
\label{triangcluster}
\end{figure}

When $\beta=\pi/2$, considering $\phi^{\sta}$ and $\phi^{\tri}$ as real films, the graphics of the reference 3-cell is equivalent to a symmetric three bubble cluster (figure \ref{triangcluster}-a). The absolute value of (\ref{epsbase}) gives its relative energy ($E/\sqrt{A}$):
$$\epsilon_3=|\epsilon_{\pi/2}^0|=(4\sqrt{3}+6\pi)^{1/2}\simeq 5.07718.$$
Applying (\ref{moeb_transf}) to the reference produces a general three bubble cluster with constant surface tension (figure \ref{triangcluster}-b).

\section{Area and energy of a general 3-cell}\label{sec_geo_quan}
To calculate the relative excess energy of a 3-cell $\widetilde{\mathcal{C}}(s,q,\beta)$, we need to calculate the star and triangle perimeters, its area and the edge curvatures.

Let us start with the edge radii.
Using  equations (\ref{trasf_cer_cer}) and (\ref{trasf_lin_cer}), they
are, for $j=1,2,3$,
\bea
\label{rtilde-int}
\tilde{r}^{\sta}_j&=&\frac{1-|s|^2}{2\,|s|\sin(\theta_j - \theta_s)}\,,\\
\tilde{r}^{\tri}_j&=&\frac{6\left(1 - |s|^2\right)\,q\sin\beta}
{4|3^{\frac{1}{2}}\sin\beta-q\bar{s}e^{i\theta_j}
  \cos(\beta+\frac{\pi}{6})|^2-3q^2 |s|^2},\quad
\label{rtilde-ext}
\eea
where $\theta_s=\arg(s)$.
As anticipated (sec. \ref{refer_3_cell}), the star radii $\tilde{r}^{\sta}=\tilde{r}^{\sta}(s)$ don't depend on $q$.
The radii inverse, that is, the curvatures, verify the equilibrium conditions (\ref{uplateauk}), and the following equations:
\bea
\label{sumKQ_in}
K_{\sta}^2(|s|)&\equiv&\sum_{j=1}^3 \frac{1}{(\tilde{r}^{\sta}_j
)^2} = 6\left(\frac{|s|}{1-|s|^2}\right)^2,\\
K_{\tri}^2(|s|,\beta)&\equiv&\sum_{j=1}^{3}\frac{1}{(\tilde{r}^{\tri}_j
)^2} =
\nonumber
\frac{8 |s|^2 \cos^2(\frac{\pi}{6} + \beta)}{(1-|s|^2 )^2} +\\
&+&\frac{4\left(|s|^2 q^2 \cos(\frac{\pi}{6}-\beta)-3\sin\beta\right)^2}{3(1-|s|^2 )^2\,q^2}.
\label{sumKQ_ex}
\eea
Thus, the sum of the squared curvatures, in the star or in the triangle, doesn't depend on the argument of $s$. The only relevant parameter left for the star sum in eq. (\ref{sumKQ_in}) is $|s|$.

\subsection{Perimeters}\label{subsec_peri}
\begin{figure}[!ht]\centering
\includegraphics[width=.8\columnwidth]{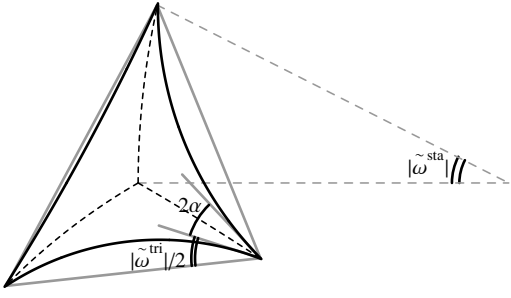}
\caption{Star and triangle for $\beta=-\pi/10$. Summing the internal angles of the (solid) grey triangle gives (\ref{sum_ang}).}
\label{somma_angoli}
\end{figure}

The value of the internal perimeter $\tilde{L}^{\sta}(s,q)$ follows from a straightforward calculation detailed in appendix \ref{app_b}.  
\ben\label{lun_int_trasf}
\widetilde{L}^{\sta}(s,q)=\sum_{j = 1}^3\,\tilde{l}^{\sta}_j=\sum_{j = 1}^3 \tilde{r}^{\sta}_j(s)\,\tilde{\omega}^{\sta}_j(s,q)\,,
\een
where
\bea
\nonumber
\tilde{\omega}^{\sta}_j(s,q)&=&2\arctan{\left(\frac{3^{-1/2} q\,|s|
-\cos(\theta_j-\theta_s)}{\sin(\theta_j -\theta_s)}\right)}+\\
\label{ang_tra_int}
&&~+2\arctan{\left(\frac{\cos(\theta_j - \theta_s)}{\sin(\theta_j - \theta_s)}\right)}
\eea
is the (signed) angle subtended by the internal film $\tilde{\phi}^{\sta}_j$ (fig. \ref{somma_angoli}).  
Once more, the internal perimeter $\widetilde{L}^{\sta}$ doesn't depend on $\beta$.

Analogously, the triangle perimeter is given by:
\ben\label{lun_ext_trasf}
\widetilde{L}^{\tri}_\beta(s,q)=\sum_{j=1}^3\,\tilde{l}^{\tri}_j=\sum_{j=1}^3\tilde{r}^{\tri}_j (s,q,\beta)\,
\tilde{\omega}^{\tri}_j(s,q,\beta)
\een
where $\tilde{\omega}^{\tri}_j(s,q,\beta)$ is the angle subtended by the triangle edge $\tilde{\phi}^{\tri}_j$, defined by (\ref{lun_est_final}).
The angles $\tilde{\omega}^{\tri}_j$ verify the equation
\ben\label{sum_ang}
\sum_{j=1}^3\tilde{\omega}^{\tri}_j=6\,\beta\,,
\een
consequence of the fact that the angles of the (rectilinear) triangle sum up to $\pi$ (see figure \ref{somma_angoli}). 

The limit value $q_{\max}(s,\beta)$ (sec. \ref{gene_3_cell}), where the boundary of the reference triangle meets the pole $1/\bar s$ of $f_s$, can be derived  from (\ref{rtilde-ext}). 
Indeed, fixing all the parameters ($s$ and $\beta$) but $q$, the sign of $k_j=1/\tilde{r}_j^{\tri}$ changes when the circle supporting $\phi_j^{\tri}$ meets the pole $1/\bar s$. For each $j=1,2,3$, this occurs at the positive solution $q=q_j$ of
\ben\label{zeri_rext}
k_j=1/\tilde{r}^{\tri}_j(s,q,\beta)=0\,.
\een
The explicit expression of $q_j$ is:
\bea
\nonumber q_j(s,\beta)&=&
\frac{\sqrt{3}\left[3-4\cos^2(\beta +\frac{\pi}{6})\sin^2(\theta_j - \theta _s)\right]^{1/2}}
{2|s|\cos(\beta-\frac{\pi}{6})} -\\
&&~-\frac{\sqrt{3}\cos(\beta+\frac{\pi}{6})\cos(\theta_j-\theta_s)}
{|s|\cos(\beta-\frac{\pi}{6})}\,.
\label{qj}
\eea

When $q$ increases, the pole is first met by circle portions outside the triangle. The bound  $q=q_{\max}(s,\beta)$ corresponds to the encounter with the triangle boundary in the last of the three circles:
\ben\label{r_max_eta}
q_{\max}(s,\beta)=\max_{j=1,2,3}q_j\,.
\een

For $q>q_{\max}$, the interior of the 3-cell is no more bounded and relation (\ref{sum_ang}) is no longer verified, as illustrated on fig. \ref{errelimit}.

\subsection{Area}\label{subsec_area}
The area of the 3-cell $\widetilde{\mathcal C}(s,q,\beta)$ is the sum of four parts: the rectilinear triangle based on the triangle vertices and the (signed) area of the three lenses around:
\ben\label{area_trasf}
\widetilde{A}_\beta(s,q)=\widetilde{A}_{\bigtriangleup}(s,q)
+\widetilde{A}_{\cup}(s,q,\beta)
\een
with
\bea\label{area_triang}
\widetilde{A}_{\bigtriangleup}&=&2\,\tilde{r}^{\tri}_1\tilde{r}^{\tri}_2
\textstyle \sin\!\left(\frac{\tilde{\omega}_1}{2}\right)
\sin\!\left(\frac{\tilde{\omega}_2}{2}\right)
\left|\sin\left(\frac{\pi}{3}-\beta+\frac{\tilde{\omega}_3}{2}\right)\right|,\quad
\\ \label{area_cups}
\widetilde{A}_{\cup}&=&\textstyle \frac{1}{2}\sum_{j=1}^3(\tilde{r}^{\tri}_j)^2
\left(\tilde{\omega}_j-\sin\tilde{\omega}_j\right).
\eea
In these equations, $\tilde{\omega}_j$ stands for $\tilde{\omega}^{\tri}_j$.
Although  $\beta$ appears in the RHS of (\ref{area_triang}), it is clear that the area of the triangle $\widetilde{A}_{\bigtriangleup}$ doesn't depend on it. Indeed, the triangle $\bigtriangleup$ is built only on the vertices, independent of $\alpha$, and so $\beta$ (they are determined by $q$ in the reference and then mapped by $f_s$).

\subsection{Relative excess energy}\label{subsec_denexcc}

Using (\ref{lun_int_trasf}), (\ref{lun_ext_trasf}) and (\ref{area_trasf}), we obtain the relative excess energy for a general 3-cell $\widetilde{\mathcal C}(s,q,\beta)$:
\ben\label{eps_trasf}
\epsilon_\beta(s,q)=\frac{\widetilde{L}^{\tri}_\beta(s,q)/(2\,\cos(\beta+\pi/6))
-\widetilde{L}^{\sta}(s,q)}
{\widetilde{A}_\beta(s,q)^{1/2}}\,.
\een

Before we draw the plots, let us make a change of variables
to take advantage of the dilation invariance of the relative excess energy, already noticed.
The new parametrisation is defined as follows:
\ben\label{new_coor}
\left\{\begin{array}{clll}
\rho&=q(1-|s|^2)\quad&\textrm{with}\ &\rho\ge 0\,,\\
\eta&=q\,|s| &\textrm{with}&0\le\eta\le\bar{\eta}(\beta, \theta_s)\,,\\
\theta_s&=\arg(s),&
\end{array} \right.
\een
where $\bar\eta(\beta, \theta_s)= |s| q_{\max}(s,\beta)$ is independent of $|s|$ by (\ref{qj}). The fourth parameter, $\beta$, is unchanged. The parameter $\rho$ corresponds to a magnification of the 3-cell, so the relative excess energy doesn't depend on it.
Moreover, from the symmetry of the reference 3-cell under the group $D_3$ generated by mirrors at $\pi/3$, the relative excess energy verifies:
\ben\label{prop_period_theta}
\epsilon_{\beta}(\eta,\theta_s)=\epsilon_{\beta}(\eta, 2\pi/3\pm\theta_s).
\een

\begin{figure}[!ht]\centering
\includegraphics[width=.95\columnwidth]{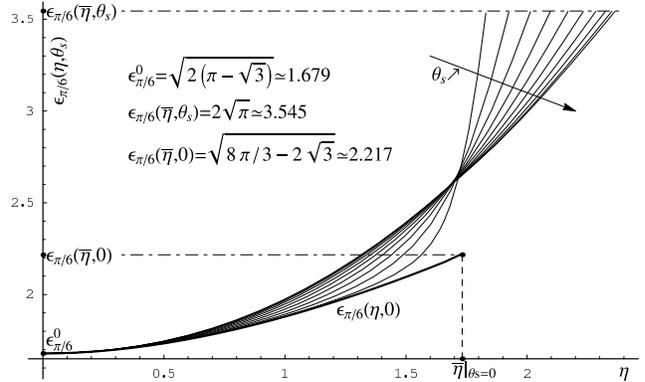}
\caption{Plot of the relative excess energy of a 3-sided bubble
  ($\beta=\pi/6$) as a function of $\eta=q|s|$ for some values of
  $0\le\theta_s\le \pi/3$. The thick solid line corresponds to $\theta_s=0$ where the system has a mirror symmetry. The dot-dashed lines point out the limit values of $\epsilon_{\pi/6}$ calculated at $\bar{\eta}$.}
\label{dis_eps_bol_rho}
\end{figure}
\begin{figure}[!ht]\centering
\includegraphics[width=.95\columnwidth]{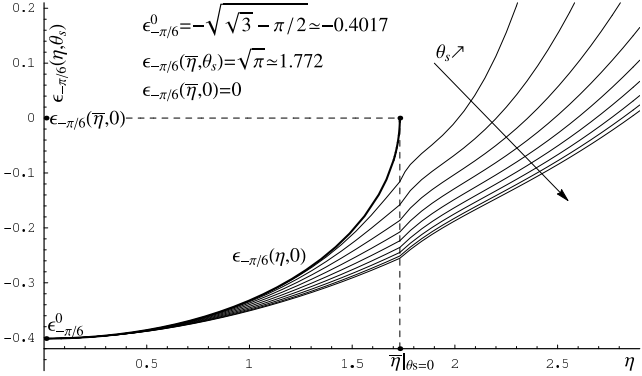}
\caption{The relative excess energy of a Plateau border ($\beta=-\pi/6$) as a function of $\eta=q|s|$ for some values of $0\le\theta_s\le \pi/3$. The solid line corresponds to $\theta_s=0$, where the border has a mirror symmetry.}
\label{dis_eps_plat_rho}
\end{figure}
Figures \ref{dis_eps_bol_rho} and \ref{dis_eps_plat_rho} show the plots of $\epsilon_{\beta}(\eta,\theta_s)$ for a 3-sided bubble and a Plateau border, respectively.
In both cases, the relative excess energy is minimal for $\eta=0$, i.e. in the symmetrical reference configuration. 

In the 3-sided bubble case, for low values of  $\eta$, $\epsilon_{\pi/6}$ increases with $\theta_s$ while, for greater $\eta$, it decreases with $\theta_s$; the change, $(\partial/\partial\theta_s)\epsilon(\pi/6, \eta, \theta_s)=0$, occurs at $\eta$ values close to $\bar{\eta}(\pi/6, 0)$.
At $\eta= \bar{\eta}(\pi/6,\theta_s$), when the area of the 3-sided bubbles diverges (sec. \ref{gene_3_cell}), $\epsilon_{\pi/6}(\bar{\eta}(\pi/6,\theta_s),\theta_s)$ is constant as a function of $\theta_s$, meaning that all the curves in fig. \ref{dis_eps_bol_rho} end at the same ordinate.

In the Plateau border case ($\beta=-\pi/6$), $\epsilon_{-\pi/6}$ is a non increasing function of $\theta_s$:
$(\partial/\partial{\theta_s})\epsilon_{-\pi/6}\le 0$. For any $\theta_s$, at $\eta=\bar{\eta}(0)$, the triangle vertices stay on a straight line and the area of the rectilinear triangle is zero: $\widetilde{A}_{\bigtriangleup}=0$.
This implies that at $\eta=\bar\eta(0)$, the derivative $(\partial/\partial{\eta})\epsilon_{-\pi/6}$ has a singularity, corresponding to the cuspidal points of $\epsilon_{-\pi/6}(\eta,\theta_s)$ in figure \ref{dis_eps_plat_rho}.
Again, in this case, at $\eta=\bar{\eta}(-\pi/6,\theta_s$) and $\theta_s\ne 0$, when the area of the Plateau border diverges (sec. \ref{gene_3_cell}), the relative excess energy doesn't depend on $\theta_s$: $\epsilon_{-\pi/6}(\bar{\eta},\theta_s)=1.772\ldots$ (outside the figure).

\subsection{Pressure as coordinates}\label{subsec_Pcoord}

Solving equations (\ref{rtilde-ext}) and inserting into the formulae of sec. \ref{subsec_peri}-\ref{subsec_denexcc} yields the relative excess energy as a function of the side radii $\tilde{r}^{\tri}_j$, $j=1,2,3$. By Laplace's equation (\ref{ulaplace}),  the curvatures are proportional to the pressure differences.
If $P_0$ denotes the pressure in the 3-cell and $P_j$ the pressure in the neighbours $j=1,2,3$, we can  analyse $\epsilon_\beta$ as a function of the pressures.

By the dilation invariance of $\epsilon_{\beta}$, we can set the absolute value of one of the sides radii to 1 without loss of generality.
So, fixing $P_0=0$ and $\tilde{r}^{\tri}_1=\mathrm{sign}(\beta)$, (\ref{ulaplace}) implies
\ben\left\{
\begin{array}{l}
{\displaystyle P_1=-\frac{\mathrm{sign}(\beta)}{2\, \cos(\beta+\pi/6)} }\,,\\[9pt]
{\displaystyle P_2=-\frac{1}{2\, \tilde{r}^{\tri}_2 \cos(\beta+\pi/6)} }\,,\\[9pt]
{\displaystyle P_3=-\frac{1}{2\, \tilde{r}^{\tri}_3 \cos(\beta+\pi/6)} }\,.
\end{array}\right.
\een

\begin{figure}[!ht]\centering
\includegraphics[width=1\columnwidth]{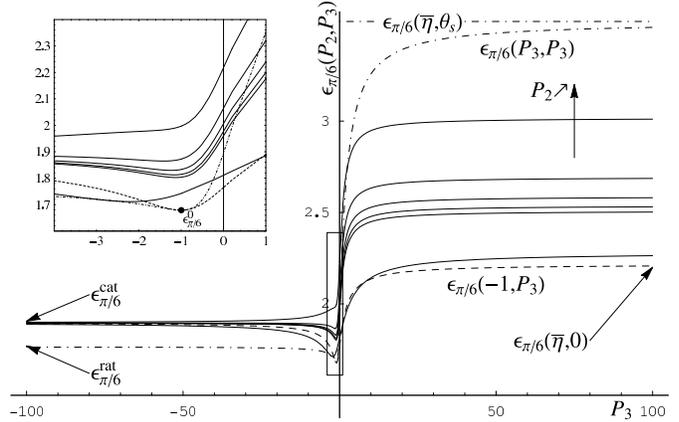}
\caption{Standard bubble case, $\beta=\pi/6$ and $P_1=-1$. Plot of $\epsilon_{\pi/6}$ versus the pressure $P_3$ for different values of $P_2$.}
\label{graf_boll_press}
\end{figure}

\begin{figure}[!ht]\centering
\includegraphics[width=1\columnwidth]{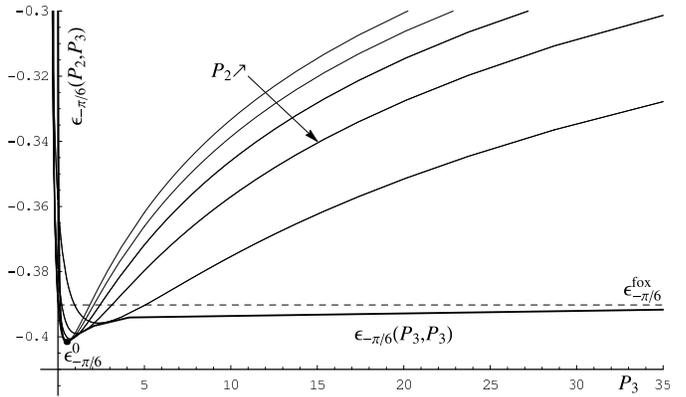}
\caption{Plateau border, $\beta=-\pi/6$ and $P_1=-1/2$.  $\epsilon_{-\pi/6}$ is plotted with respect to  $P_3$ for different values of $P_2$.}
\label{graf_plat_press}
\end{figure}

In figures \ref{graf_boll_press} and \ref{graf_plat_press} ($\beta=\pi/6$ and $\beta=-\pi/6$, respectively), we have drawn the parametric plots of the relative excess energy as a function of $P_3$ for different values of $P_2$. 

\begin{figure}[!ht]\centering
\includegraphics[width=1\columnwidth]{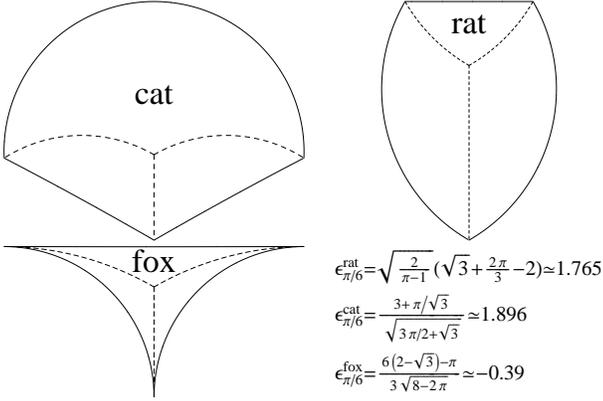}
\caption{Limit configurations found when one or two of the pressure differences with the neighbours go to infinity.  "Rat" and "cat" describe 3-sided bubbles, while "fox" appears for Plateau borders.}
\label{conf_limit}
\end{figure}

In figure \ref{graf_boll_press}, the limit of $\epsilon_{\pi/6}(P_2,P_3)$ when $P_3\to-\infty$ corresponds to configurations of a non-symmetrical cell when the area of the 3-sided bubble goes to zero. $P_2=P_3\to-\infty$ gives the configuration "rat", whereas in all the other cases, $P_3\to-\infty$ at fixed $P_2$, the bubble goes to the configuration "cat" (fig. \ref{conf_limit}). 

In the Plateau border case (figure \ref{graf_plat_press}), when $P_2=P_3\to+\infty$, the border goes to a configuration "fox" (fig. \ref{conf_limit}), where one of the edges is straight. In the other cases, $\epsilon_{-\pi/6}$ goes to zero. Obviously, in all cases, bubble or border, equal pressure ($P_1=P_2=P_3$) gives $\epsilon_\beta(P_1,P_1)=\epsilon_\beta^0$.

\section{Decorating a fixed star}\label{sec_foam_fix}
In this section we consider an ideal dry 2D foam at equilibrium and we ask how much the energy of the foam varies if we replace one vertex by a Plateau border or a triangular bubble.

\begin{figure}[!ht]\centering
\includegraphics[width=.6\columnwidth]{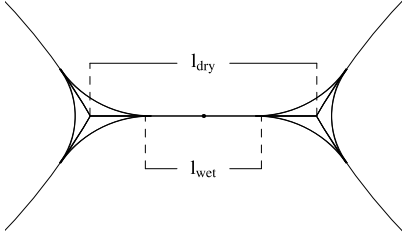}
\caption{The distance between two vertices in dry a foam, $l_{dry}$, is larger than in a wet foam, $l_{wet}$.}
\label{Ti_wet_dry}
\end{figure}

This question is important in foam rheology \cite{princen83,hutzler_05,wea_web_03}. 
If the foam is not dry, the film length is reduced by the presence of the Plateau borders (see figure \ref{Ti_wet_dry}), favouring the switch.
Preliminary calculations on small samples show that the presence of liquid Plateau borders reduces the T1 energy barrier, whereas the presence of triangular bubbles leads to locally more stable quadrangles, thus increasing an effective T1 barrier.
Qualitatively, as compared to a bare foam, the net result is a decrease of the yield strain and stress when the foam is decorated by Plateau borders, but an increase when the decoration is by triangular bubbles.

More extensive investigations are in progress to get the full rheology.
Here, we show that the energy contribution due to decoration can be put into a simple tractable form.

Varying the parameter $q$ and fixing $s$, we can vary the decorating triangle area, $\widetilde{A}_\beta(s,q)$, keeping constant the star radii $\tilde{r}^{\sta}_j$. 
As said in the introduction, the size of the 3-cells in real foams will, in general, be smaller than the typical size of other bubbles. In this approximation, we can write the relative excess energy as a function of the triangle area. To do so, we expand  $\tilde{L}^{\tri}$, $\tilde{L}^{\sta}$ and $\widetilde{A}_\beta(s,q)$  with respect to  $q$, near $q=0$:
\bea\label{serie_L_ex}
\tilde{L}^{\tri}(s,q)&=&(1-|s|^2)\left(L_1^{\tri}q+L_3^{\tri}|s|^2q^3\right)+O(q^4),\qquad\\
\label{serie_L_in}
\tilde{L}^{\sta}(s,q)&=&(1-|s|^2)\left(L_1^{\sta}q+L_3^{\sta}|s|^2q^3\right)+O(q^4),\\
\label{serie_A}
\widetilde{A}_\beta(s,q)&=&(1-|s|^2)^2\left(A_2 q^2+A_4 q^4\right)+O(q^5).
\eea
The coefficients of $q^n$, which depend only on $\beta$, are given in appendix \ref{app_c}.
Using the expansion to second order in (\ref{serie_A}) to explicit $q$ as a function of $\widetilde{A}_\beta$ and replacing this in the expression (\ref{eps_trasf}) of $\epsilon_{\beta}(s,q)$, we find:
\ben\label{appr_eps}
\epsilon_{\beta}\left(s,\widetilde{A}_\beta\right)=\epsilon^0_\beta+\Gamma_{\beta}\,
K_{\sta}^2(|s|)\,\widetilde{A}_\beta+O(\tilde{A}_\beta^{3/2})\,,
\een
where $\epsilon^0_\beta$ and $K_{\sta}^2$ are given by (\ref{epsbase}) and (\ref{sumKQ_in}), respectively. 
So the excess energy defined by Teixeira and Fortes \cite{Fortes05} is, for small areas:
\ben\label{appr_E}
E_\beta\left(s,\widetilde{A}_\beta\right)=\epsilon^0_\beta\,\widetilde{A}_\beta^{1/2}+\Gamma_{\beta}K_{\sta}^2(|s|)\,\widetilde{A}_\beta^{3/2}+O(\tilde{A}_\beta^2).
\een
The factor $\Gamma_{\beta}$ in (\ref{appr_eps}), (\ref{appr_E}) depends only on $\beta$ and its expression is: 
\ben
\Gamma_{\beta}=\frac{1}{6 A_2^{3/2}}\left\lbrack\frac{L_3^{\tri}-L_1^{\tri} \frac{A_4}{2A_2}}{2 \cos(\beta+\frac{\pi}{6})}+L_1^{\sta} \frac{A_4}{2A_2}-L_3^{\sta}\right\rbrack.
\een

Given a fixed star in a 2D dry foam, the relative excess energy produced by decoration is, for small areas, linear in $\widetilde{A}$. The angular coefficient involves the numerical factor $\Gamma_{\beta}$, depending on the type of the decorating 3-cell, and the star sum of the square curvatures $K_{\sta}^2$.       

In particular, the relative excess energy depends on $|s|$ only through the product of the triangle area and the  mean square curvature of the star:
\ben
\epsilon_\beta\left(s,\widetilde{A}_\beta\right)
=\epsilon_\beta\left(\widetilde{A}_\beta\cdot K_{\sta}^2(|s|),\theta_s\right).
\een
In terms of the coordinates (\ref{new_coor}), $\widetilde{A}_\beta\cdot K_{\sta}^2$ does not depend on $\rho$. 
The argument $\theta_s$ appears only in higher order terms of the expansions of $\epsilon_\beta$. In (\ref{appr_eps}), the $\widetilde{A}_\beta^{3/2}$ term is proportional to $K_{\sta}^{3/2}\!\cos(3\,\theta_s)$.

\begin{figure}[!ht]\centering
\includegraphics[width=.95\columnwidth]{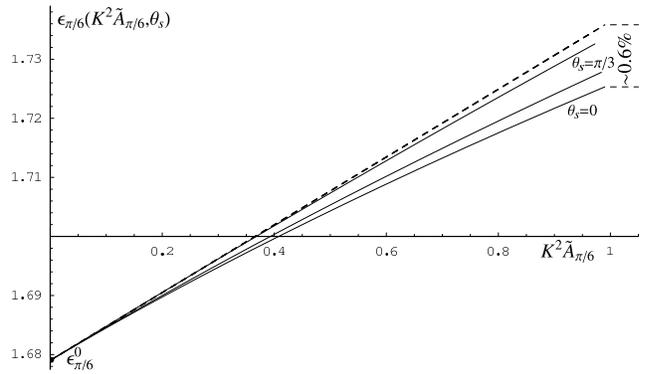}
\caption{The relative excess energy versus $\tilde{A}_\beta K_{\sta}^2$ for different values of $\theta_s$ at $\beta=\pi/6$. Dashed, the linear approximation of $\epsilon_\beta$. For $\tilde{A}_\beta K_{\sta}^2<1$, the error is smaller than 0.6\%.}
\label{appr_linea}
\end{figure}

Figure \ref{appr_linea} shows the plot of $\epsilon_\beta$ versus $\widetilde{A}_\beta K_{\sta}^2$ in the case of a three-sided bubble.  

The positive quantity $K_{\sta}^2$, depending only on $|s|$, has the dimension of an inverse area. For a symmetrical star, $K_{\sta}^2(0)=0$, and it increases with $|s|$.
Let us remark that, by (\ref{uplateauk}), we can write:
\ben
K_{\sta}^2=2\left\lbrack(k_1)^2-k_1 k_2+ (k_2)^2\right\rbrack,
\een
where for simplicity we call $k_j=1/\tilde{r}^{\sta}_j$. This quantity appears when 2D dry foams are studied using  projective geometry and local coordinates associated to any vertex \cite{MancioTesi}.

\section{The flower problem}
\begin{figure}[!ht]\centering
a)\includegraphics[width=.28\columnwidth]{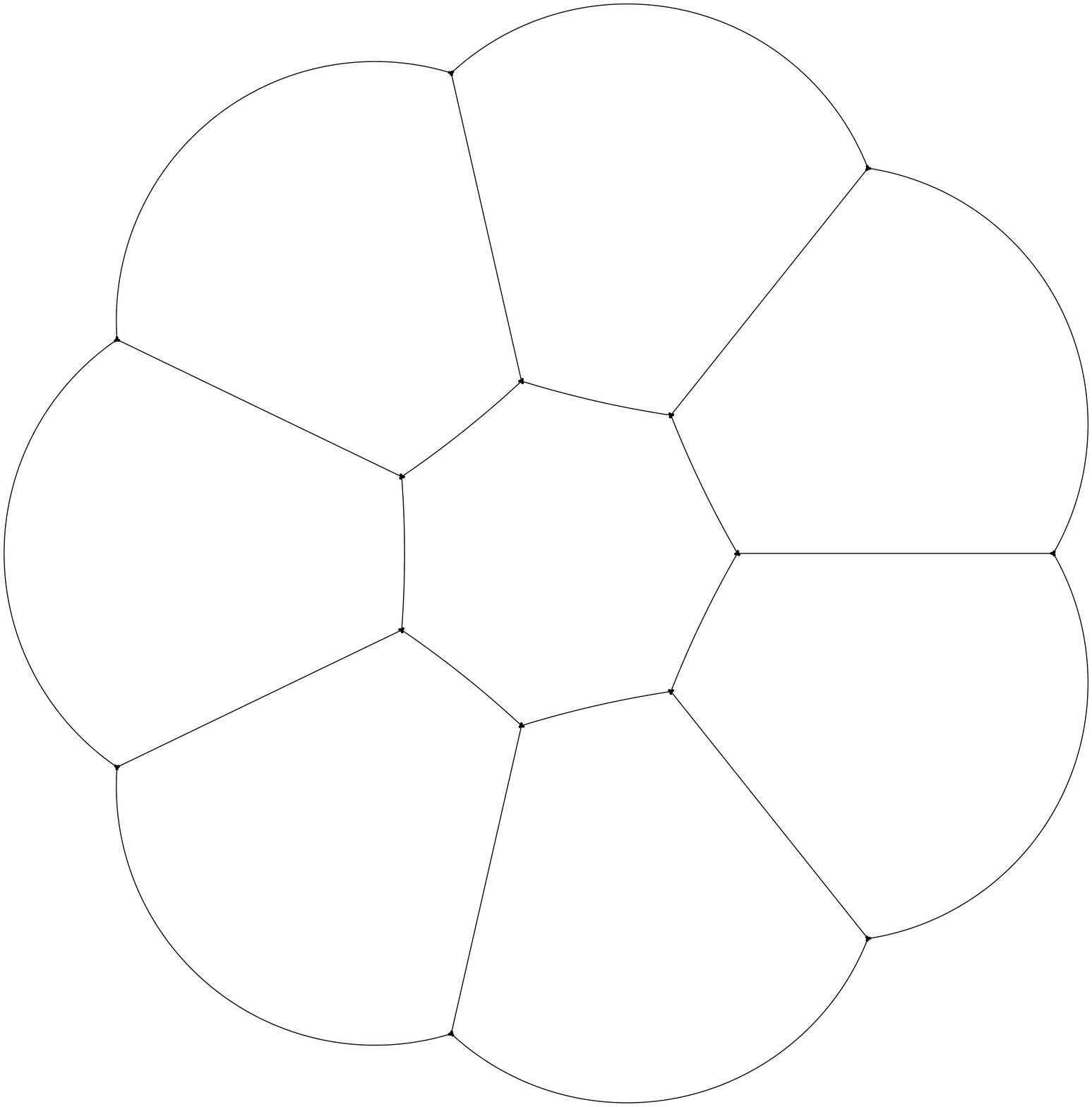}
b)\includegraphics[width=.28\columnwidth]{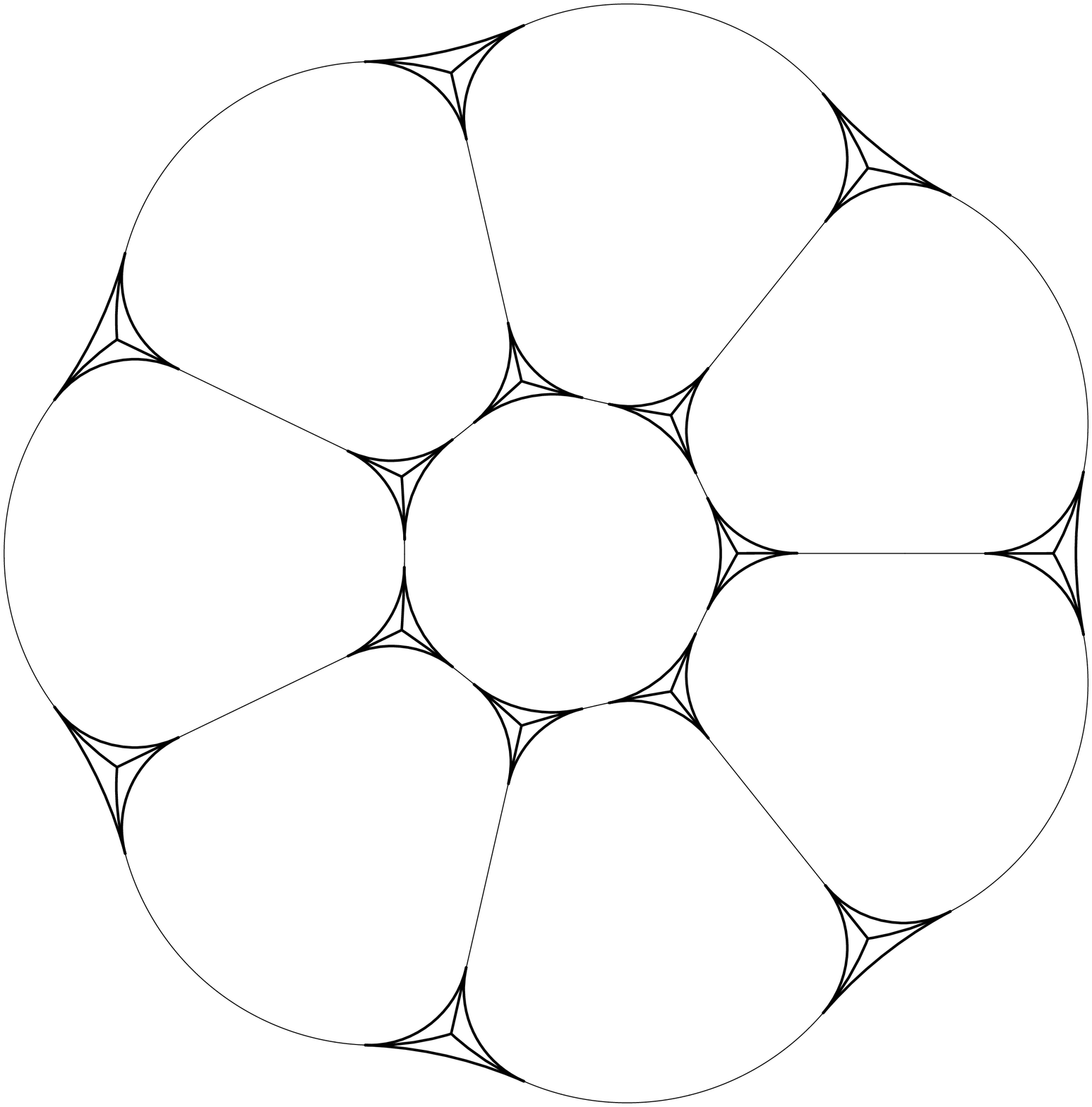}
c)\includegraphics[width=.28\columnwidth]{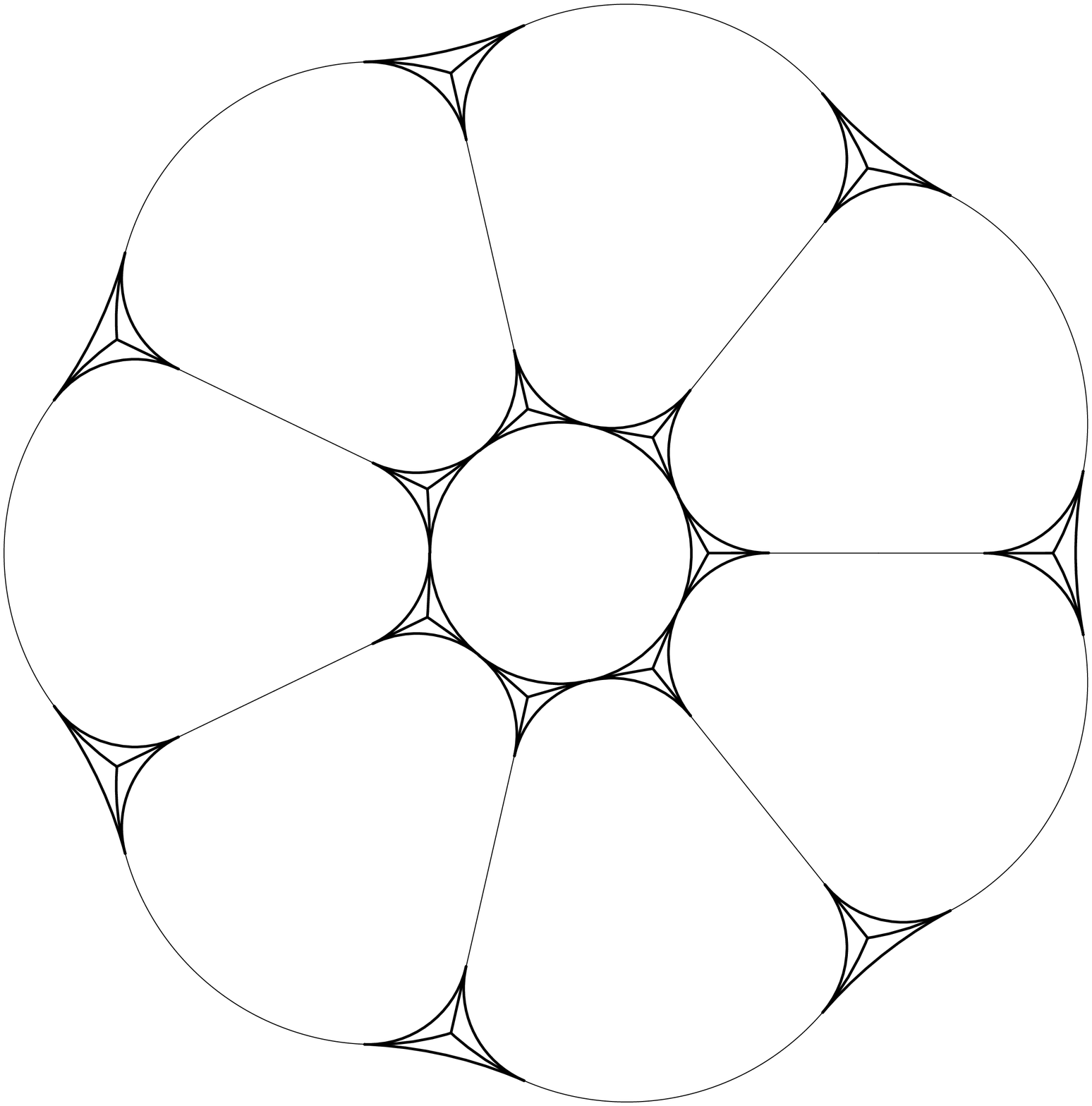}
\caption{Flower problem. a) The flower cluster without liquid. b) The vertices are decorated by Plateau borders. c) The limit where some Plateau borders touch each other.}
\label{petalo_p_figure}
\end{figure}

The flower problem, introduced by Weaire {\it et al} \cite{Wea_Cox_Gran_flower_02}, consists of a dry cluster of $n$ equal-sized bubbles surrounding a central bubble of area $A_c$ (fig. \ref{petalo_p_figure}-a).  By energy minimisation using Surface Evolver, the authors computed the minimal energy configurations as a function of $A_c$.
They found that, for $n>6$, the symmetrical configuration is minimal when $A_c$ is bigger than a critical area $A_c^*$. When $A_c<A_c^*$ the symmetry of the cluster is reduced and the energy landscape  contains multiple, but energetically equal, minima  (as shown in \cite{Cox_flower_03}). The degeneracy follows from the spontaneously broken $n$-fold symmetry.
The value of $A_c^*$ corresponds to a pressure of the central bubble equal to the pressure outside the cluster.

In Cox {\it et al}'s experimental set-up \cite{Cox_flower_03}, the bubbles were trapped between a glass plate and the top of the liquid. The area of the central bubble was slowly decreased but  the predicted symmetry breaking was not observed. Before the critical area, at some $A_c^{**}>A_c^{*}$,  a T1 instability occurred and a bubble was ejected from the cluster. The authors explain the ejection by the presence a small amount of liquid, which provokes the T1, and by some effects due to the third dimension.

Considering only the effect of the liquid, we estimate its influence on the occurrence of topological changes and the petal ejection. In the dry 2D model, when a film length shrinks to zero, the unstable 4-fold vertex relaxes to a lower energy configuration where an edge joins two 3-fold vertices.

Now, with a little liquid at the vertices, the film length is reduced by the presence of the Plateau borders, so that the topological change is likely to occur earlier. In the flower,  because of the Plateau borders, the film length may vanish before the critical area $A_c^{*}$ is reached. This is precisely confirmed by our calculations.

We have decorated the flower vertices by Plateau borders at constant pressure (fig. \ref{petalo_p_figure}-b). The borders don't have the same area. Moreover by the presence of a liquid reservoir in the experimental set-up \cite{Cox_flower_03},  the liquid fraction is not fixed. Then, we calculated the value  of the central area,  $A_{\lim}$, at which Plateau borders touch each other (fig. \ref{petalo_p_figure}-c), when the ejection is conjectured occur. 

As we didn't know the experimental liquid pressure, we chose the value, of the pressure in the Plateau borders, that agreed best with the experimental points. This value is $P_0/\gamma_i=-0.390\pm 0.015\,m^{-1}$, provided the external pressure is set to zero.   

\begin{figure}[!ht]\centering
\includegraphics[width=.95\columnwidth]{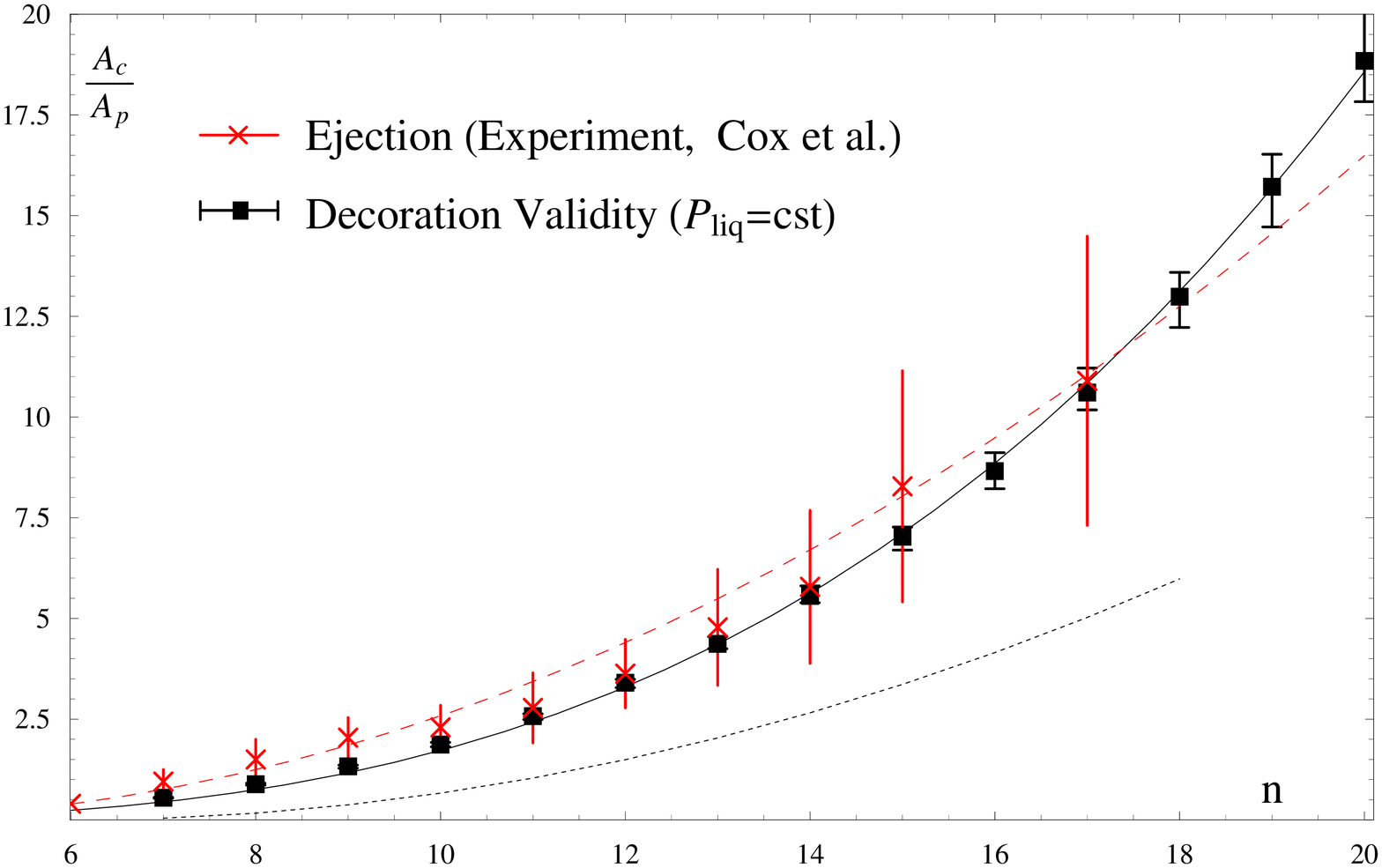}
\caption{
Critical area of the central bubble, normalised by the petal area, versus the number of petals $n$.  
Our calculations are compared to the experimental data from \cite{Cox_flower_03}. 
Squares: the calculated values at which two Plateau borders touch each other when the central area decreases at constant liquid pressure ($A_{\lim}$); these values are close to the solid curve of equation $A_c/A_p = 0.003(n-1.75)^3$.
Crosses: Cox {\it et al}'s experiment \cite{Cox_flower_03}, the mean critical value at which the central bubble is ejected when its area decreases ($A_c^{**}$). The dashed line is the quadratic fit from \cite{Cox_flower_03}, as described in text.
Dotted, the theoretical curve where the dry flower cluster would undergo symmetry reduction \cite{Wea_Cox_Gran_flower_02}.} 
\label{petalo_p_cst_a_n}
\end{figure}
Figure \ref{petalo_p_cst_a_n} compares the values of $A_{\lim}$ and the experimental $A_c^{**}$ \cite{Cox_flower_03} as functions of $n$.
Our values are much closer to the experimental data than those of the symmetry reduction criterion. The calculated values of $A_{\lim}$  fit very well to a cubic power law:
\ben
A_{\lim}= 0.003 (n-1.75)^3A_p\,.
\een
This result  disagrees with the quadratic law in \cite{Cox_flower_03}. This quadratic law was found by adjusting the central area so that the energy of the symmetric cluster and that of the ejected configuration be equal, as calculated by Surface Evolver \cite{SurfEvo}. 

In figure \ref{petalo_p_cst_a_n}, the small, but systematic, difference between the experimental ejection area and $A_{\lim}$ may be explained by considering that, experimentally, the size of the bubbles can not be exactly the same, thereby reducing the shortest film length and triggering T1 a bit earlier than in an exactly symmetric model.

\begin{figure}[!ht]\centering
\includegraphics[width=.95\columnwidth]{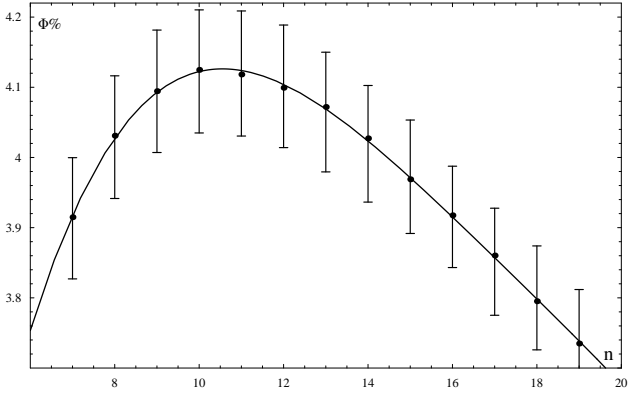}
\caption{Values of the liquid fraction obtained at the limit decoration versus the petal number $n$ (figure \ref{petalo_p_cst_a_n}). }
\label{petalo_phi_n}
\end{figure}

Figure \ref{petalo_phi_n} shows the predicted liquid fraction at $A_{\lim}$ as a function of $n$. Here, the liquid fraction is defined as the ratio of the liquid to total (enclosed gas+liquid) area~\footnote{This $\Phi$ may differ from global liquid fractions in not including any gas or area outside the cluster.}.
At constant pressure, the liquid volume fraction has to vary. The experiments appear to have been performed at about $4\%$ liquid fraction\footnote{This is an area fraction in the 2D model. Specifying the physical liquid volume or fraction is more delicate, if relevant at all, due to the finite thickness in the third dimension and the liquid borders along the plate(s).}.

\section{Conclusions}

To summarise, in a 2D equilibrated foam, star-triangle equivalence permits to replace dry three-fold vertices by three-sided cells and \textit{vice versa}. The general structure of a decorated star was obtained by first defining a two real parameters reference 3-cell, then applying a one complex parameter Moebius transformation. This way allowed us to give to the Plateau border and 3-sided bubble problems a unified description. We calculated all the geometrical and physical quantities of interest: area, perimeter and energy.
Beside Plateau borders and standard three-sided bubbles, the method also applies to triple partitions of the disk and the three-bubble cluster.

The difference between the decorated (triangle) and bare (star) vertex is quantified by the relative excess energy: 
$$\epsilon_\beta=E_\beta/\sqrt{A}\,,$$
the energy difference normalised by the square root of the triangle area.
This dimensionless quantity, depending on only three parameters, serves as a shape discriminant. At given area, the most symmetric configuration minimises $\epsilon_\beta$. 

Fixing the equilibrated star, the relative excess energy of 3-cells is affine in $A$ with a good approximation:
$$\epsilon_\beta(A)\simeq\epsilon^0+b A,$$
where $b$ depends on the form of the decorated films.
The slope $b$  is, up to a positive factor, the  mean square curvature of the star ($\tilde{K}_{\sta}^2$); being zero in the 3-fold symmetric case, $b$ measures the deviation from perfect symmetry.

As an application to the flower problem, we calculated, for various petal numbers and including liquid at the vertices, the limit value of the central area when two Plateau borders get into contact. Tuning the liquid pressure to fit the experimental data, we obtained a nice agreement of the calculated threshold with the experiment. We predict an experimental liquid fraction of about $4\%$. 

The calculated threshold is weakly sensitive to the liquid pressure. 
Very likely, the small systematic difference between the calculated and experimental data comes from a slight discrepancy of the experimental  bubble size with exact $n$-fold symmetry. The calculated limit central area, as a function of the number of petals $n$, fits very well a cubic law.   

\subsection*{Acknowledgement}
\begin{acknowledgement}
We would like to thank N. Rivier and S. Cox for discussions about the description of a 3-sided cell and the flower problem, respectively.

\end{acknowledgement}

\section*{Appendixes}
\begin{appendix}

\section{Moebius transformations}\label{app_a}
A Moebius transformation or homography, $\tilde{z}=f(z)$, is an automorphism of the augmented complex plane ($\mathbb{C}^*\equiv \mathbb{C} \cup \{\infty\}$), $f:\mathbb{C}^*\to\mathbb{C}^*$ \cite{conformMappinBook,Dubrovin}, defined by:  
\ben\label{moeb_gen_trans}
\tilde{z}=f(z)=\frac{a\,z+b}{c\,z+d}\,,\quad \textrm{with}\,\, a d-b c = 1\,.
\een 
A Moebius transformation is a combination of translations, rotations, dilatation and inversions \cite{conformMappinBook}. 

Beside preserving angles (conformality), these transformations map circles to circles; in particular any circle passing through the inversion point, $v^*=-d/c$, is sent to a straight line (circle through infinity).

The circle of parameters $(z_0=0,r)\in\mathbb{C}\times\mathbb{R}$ (centre and radius) is mapped by (\ref{moeb_gen_trans}) to the circle of parameters:
\ben\label{trasf_cer_cer}
\left(\tilde{z}_0,\tilde{r}\right)=
\left(
\frac{a\,\bar{c}\,r^2-b\,\bar{d}}{|c|^2\,r^2-|d|^2},
\frac{|a\,d-b\,c|\,r}{|r^2\,|c|^2-|d|^2|}
\right),
\een 
where $\bar{x}$ stands for the complex conjugate of $x$.

Similarly, $f$ maps the straight line of parametric equation $z=z_0+m t$, $t\in U\subset\mathbb{R}$, with $z_0=0\,,m\in\mathbb{C}$, to the circle of parameters:
\ben\label{trasf_lin_cer}
\left(\tilde{z}_0,\tilde{r}\right)=
\left(
\frac{2\,a\,\mathrm{Im}(m\,c\,\bar{d})+i\,\bar{c}\,\bar{m}}{2\,c\, \mathrm{Im}(c\,m\,\bar{d}) },
\frac{|a\,d-b\,c|\,|m|}{|2\, \mathrm{Im}(c\,m\,\bar{d})| }
\right).
\een 
When $z_0\neq0$, equations (\ref{trasf_cer_cer}) and (\ref{trasf_lin_cer}) keep their validity with the substitution: $b\to b+az_0$ and $d\to d+cz_0$.

\section{3-cell perimeters}\label{app_b}
Let us consider an edge defined by its parametrisation $\phi(u)$, $u\in\lbrack u_0,u_1\rbrack$. In the transformed plane, the length of $\tilde{\phi}=f_s\circ\phi$, image  of $\phi$ by $f_s$, is
\bea\label{lun_film_gen}
\nonumber\tilde{l}&=&\int_{u_0}^{u_1} |\dot{\tilde{\phi}}(u)|\,\d u=
\int_{u_0}^{u_1}\left|\frac{\d{f_s}}{\d{z}}(\phi(u))\right| |\dot{\phi}(u)|\,\d u=\\
&=&(1-|s|^2)\int_{u_0}^{u_1}\frac{1}{\left|1-\bar{s}\,\phi(u)\right|^2}
\,|\dot{\phi}(u)|\,\d u\,.
\eea
\subsection{Internal perimeter}\label{subsec_int_per}
For an internal film $\phi^{\sta}_j(t_1)$, substituting equation (\ref{fi_int_z}) in (\ref{lun_film_gen}) gives, after simplification:
\bea\label{lun_int_calc}
\nonumber\tilde{l}^{\sta}_j&=&(1-|s|^2)\int_{0}^{\frac{r}{\sqrt{3}}}\frac{\d u}
{1-2\,u|s|\cos(\theta_s-\theta_j)+u^2\,|s|^2}\\
&=&\tilde{r}^{\sta}_j(s)\,\tilde{\omega}^{\sta}_j(s,q)\,,
\eea
where $\tilde{\omega}^{\sta}_j(s,q)$ is given in (\ref{ang_tra_int}).
\subsection{Triangle perimeter}\label{subsec_ext_per}
For a triangle edge $\phi^{\tri}_j(t_2)$, (\ref{lun_film_gen}) can be reduced to an integral of the form:
\ben\label{lun_ext_calc}
\tilde{l}^{\tri}_j=\frac{q\,(1-|s|^2)}{2\,|\sin{\beta}\,|}
\int_{-1}^{1}\frac{\beta\d u}
{a_0^j+a_1^j\cos{u\beta}+a_2^j\sin{u\beta}}\,,
\een
where the constants in the integral are:
\bea
\nonumber a_0^j&=&1+\frac{|s|\,q}{\sqrt{3}}\cos(\theta_j-\theta_s)(1-\sqrt{3}\cot{\beta})+\\
&&\qquad ~+\frac{|s|^2\,q^2}{12 \sin^2\beta}(5+2\,\cos(2\beta+\pi/3))\,,\\
a_1^j&=&\frac{|s|\,q}{\sin{\beta}}\cos(\theta_j-\theta_s)+
\frac{|s|^2\,q^2 (1-\sqrt{3}\cot{\beta})}{2\sqrt{3}\, \sin{\beta}}\,,\\
a_2^j&=&-|s|\,q\,\frac{\sin(\theta_j-\theta_s)}{\sin{\beta}}\,;
\eea
they verify
\ben\label{somm_coeff}
(a_0^j)^2-(a_1^j)^2-(a_2^j)^2=\frac{q^2\,(1-|s|^2)^2}{4\,(\tilde{r}^{\tri}_j)^2 \sin^2\beta}>0\,.
\een
Let $\Delta$ 
be the square root of the previous expression.
Then the solution of the integral (\ref{lun_ext_calc}), verifying (\ref{somm_coeff}),  is \cite{table_integr}: 
$$\tilde{l}_j^{\tri}=\frac{q\,(1-|s|^2)}{\Delta|\sin{\beta}\,|}\left.\!\arctan\left(\frac{(a_0^j-a_1^j)\tan(\frac{u}{2})+a_2^j}
{\Delta}\right)\right|_{-\beta}^{+\beta}\,.
$$ 
Taking, for $\Delta$, the root of the second member of (\ref{somm_coeff}) gives
\bea
\tilde{l}_j^{\tri}&=&\tilde{r}^{\tri}_j(s,q,\beta)\, \tilde{\omega}^{\tri}_j(s,q,\beta)\label{lun_w_final},\\
\tilde{\omega}^{\tri}_j&=&2\left. \arctan\!\left({\textstyle 2\tilde{r}^{\tri}_j \sin{\beta}\frac{(a_0^j-a_1^j)\tan(u/2)+a_2^j}{q\,(1-|s|^2)} }\right)\right |_{-\beta}^{+\beta}.\quad
\label{lun_est_final}
\eea 
Equations (\ref{lun_w_final}, \ref{lun_est_final}) also define the angle $\tilde{\omega}^{\tri}_j$ subtended by the  edge $\tilde{\phi}_j^{\tri}$.

\section{Series coefficients of $\epsilon_\beta$}\label{app_c}
The coefficients in the expressions (\ref{serie_L_ex}), (\ref{serie_L_in}) and (\ref{serie_A}) are
\begin{eqnarray*}
L_1^{\sta}&=&\sqrt{3}\,,\qquad L_3^{\sta}=\frac{1}{3\sqrt{3}}\,,\qquad L_1^{\tri}=\frac{\beta}{\sin{\beta}}\,,\\
L_3^{\tri}&=&\frac{\sqrt{3}+5\beta-2\sqrt{3}\sin(2\beta+\frac{\pi}{6})-2\beta\sin(2\beta-\frac{\pi}{6})}{4 \sin^3(\beta)}\,,\\
A_2&=&\frac{1}{4 \sin\beta}\left(\frac{3 \beta}{\sin\beta}-\frac{2}{\cos(\beta+\pi/6)}\right),\\
A_4&=&\frac{1}{8}\left[4(\sqrt{3}-\beta) - (9+4\sqrt{3}\,\beta)\cot\beta\right]\csc^2\beta +\\
&& \hspace{15mm} 
~+\frac{9}{8}\,\beta\csc^4\beta - \frac{1}{4\sqrt{3}}(1+\sqrt{3}\cot\beta)\,.
\end{eqnarray*}
The zeroth and second order (in $q$) coefficients of $\tilde{L}^{\tri}$ and $\tilde{L}^{\sta}$ are zero at $q=0$. In the same way $A_0$, $A_1$ and $A_3$ are zero.  

\end{appendix}



\end{document}